\documentclass[twocolumn]{aastex7}
\usepackage{graphicx}
\usepackage{txfonts}
\usepackage{hyperref}
\usepackage{natbib}
\begin{document} 
   
\title{Young Stellar Objects in the Carina nebula: Near-Infrared variability and spectroscopy.}

\author[orcid=0000-0002-5936-7718,sname='Borissova']{Jura Borissova}
\affiliation{Instituto de F\'isica y Astronom\'ia, Universidad de Valpara\'iso, Av. Gran Breta\~na 1111, Playa Ancha, Casilla 5030, Chile}
\affiliation{Millennium Institute of Astrophysics (MAS), Nuncio Monseñor Sótero Sanz 100, Providencia, Santiago, Chile}
\email[show]{jura.borissova@uv.cl}  

\author[orcid=0000-0002-9740-9974,sname='Kurtev']{Radostin Kurtev}
\affiliation{Instituto de F\'isica y Astronom\'ia, Universidad de Valpara\'iso, Av. Gran Breta\~na 1111, Playa Ancha, Casilla 5030, Chile}
\affiliation{Millennium Institute of Astrophysics (MAS), Nuncio Monseñor Sótero Sanz 100, Providencia, Santiago, Chile}
\email[]{radostin.kurtev@uv.cl}  

\author[sname='Escobar']{Josemanuel Escobar}
\affiliation{Instituto de F\'isica y Astronom\'ia, Universidad de Valpara\'iso, Av. Gran Breta\~na 1111, Playa Ancha, Casilla 5030, Chile}
\email[]{josemanuel.escobar@alumnos.uv.cl}  

\author[orcid=0000-0003-3496-3772,sname='Alonso-Garc\'ia']{Javier Alonso-Garc\'ia}
\affiliation{Centro de Astronomía (CITEVA), Universidad de Antofagasta, Av. Angamos 601, Antofagasta, Chile}
\affiliation{Millennium  Institute of Astrophysics (MAS), Nuncio Monseñor Sótero Sanz 100, Providencia, Santiago, Chile}
\email[]{javier.alonso@uantof.cl}  

\author[sname='Medina']{Nicolas Medina}
\affiliation{Millennium  Institute of Astrophysics (MAS), Nuncio Monseñor Sótero Sanz 100, Providencia, Santiago, Chile}
\email[]{nicomedinap@gmail.com}  

\author[sname='Osses']{Javier Osses}
\affiliation{Instituto de F\'isica y Astronom\'ia, Universidad de Valpara\'iso, Av. Gran Breta\~na 1111, Playa Ancha, Casilla 5030, Chile}
\affiliation{Millennium  Institute of Astrophysics (MAS), Nuncio Monseñor Sótero Sanz 100, Providencia, Santiago, Chile}
\email[]{javier.ossesp@postgrado.uv.cl}  

\author[orcid=0000-0003-0292-4832,sname='Guo']{Zhen Guo}
\affiliation{Instituto de F\'isica y Astronom\'ia, Universidad de Valpara\'iso, Av. Gran Breta\~na 1111, Playa Ancha, Casilla 5030, Chile}
\affiliation{Millennium Institute of Astrophysics (MAS), Nuncio Monseñor Sótero Sanz 100, Providencia, Santiago, Chile}
\email[]{zhen.guo@uv.cl}  

\author[orcid=0000-0002-8872-4462,sname='Lucas']{Philip Lucas}
\affiliation{Centre for Astrophysics, University of Hertfordshire, College Lane, Hatfield, AL10 9AB, UK}
\email[]{p.w.lucas@herts.ac.uk}  

\author[orcid=0000-0002-0631-7514,sname='Kuhn']{Michael Kuhn}
\affiliation{Centre for Astrophysics, University of Hertfordshire, College Lane,  Hatfield, AL10 9AB, UK}
\email[]{m.kuhn@herts.ac.uk}  

\author[orcid=0000-0002-7064-099X,sname='Minniti']{Dante Minniti}
\affiliation{Departamento de F\'isica y Astronom\'ia, Universidad And\'es Bello, Av. Fernandez Concha 700, Las Condes, Santiago, Chile}
\affiliation{Specola Vaticana, Vatican Observatory, Castelgandolfo, V00120 Stato Citta Vaticano, Italy}
\email[]{vvvdante@gmail.com}  

\author[orcid=0000-0001-6914-7797,sname='Covey']{Kevin R. Covey}
\affiliation{Department of Physics and Astronomy, Western Washington University, Bellingham, WA, USA}
\email[]{coveyk@wwu.edu}  

\author[orcid=0000-0001-6878-8648,sname='Saito']{Roberto Saito}
\affiliation{Departamento de F\'isica, Universidade Federal de Santa Catarina, Trindade 88040-900, Florian\'opolis-SC, Brazil}
\email[]{robsaito@gmail.com}  

\author[orcid=0000-0003-3459-2270,sname='Forster']{Francisco F\"{o}rster}
\affiliation{Millennium Institute of Astrophysics (MAS), Nuncio Monseñor Sótero Sanz 100, Providencia, Santiago, Chile}
\affiliation{Departamento de Astronomía, Universidad de Chile, Casilla 36D, Santiago, Chile}
\affiliation{Data and Artificial Intelligence Initiative (IDIA), Faculty of Physical and Mathematical Sciences, Universidad de Chile, Chile}
\affiliation{Center for Mathematical Modeling, Universidad de Chile, Beauchef 851, Santiago 8370456, Chile}
\email[]{francisco.forster@gmail.com}  

 \begin{abstract}
 
We present a catalog of 652 young stellar objects (YSOs) in the Carina star-forming region. The catalog was constructed by combining near-infrared $K_S$-band variability from the VISTA Variables in the Vía Láctea eXtended survey and medium-resolution $H$-band spectroscopy from APOGEE-2, Sloan Digital Sky Survey IV (SDSS-IV). Variability analysis of 6.35 million sources identified 606 variable stars. The classification of the spectral lines by semisupervised  K-means clustering of 704 stars, refined through comparison with known catalogs in literature and visual inspection of the spectra, was performed. Combined with $K_S$ variability, the final catalog contains three groups: Emission-line YSOs, Absorption-line YSOs, and Literature/Variable-identified YSOs. Cross validation with the Gaia DR3 proper motion and distance estimates supports Carina membership for 415 sources. The statistical characterization of YSO variability demonstrated that most Carina members (78\%) exhibit variability patterns. Of these, 134 stars show emissions in their spectra, which is consistent with some accretion processes.  Analysis of fundamental stellar parameters from StarHorse and Gaia DR3 reveals typical distributions of YSOs, dominated by low-mass (1–4\(M_\odot\)), solar-metallicity stars with temperatures between 4000 and 6000K. Only a small fraction (4\%) of the sources are more massive than 4\(M_\odot\), suggesting limited ongoing  massive star formation in Carina. This well-characterized catalog also offers a robust training dataset for machine learning applications aimed at predicting YSO behavior.

\end{abstract}


\section{Introduction}

The Carina star-forming region, situated approximately 2.4 kpc away within the Carina–Sagittarius arm of the Milky Way, is one of the most massive, active, and dynamic star-forming complexes in the Galaxy. It has long served as a natural laboratory for studying recent star-formation activity \citep{Povich11, Kuhn21, {2014A&A...567A.109K},{2021ApJ...917...23K}, {2021AJ....162..282M}, {2022A&A...660A..11G}}. The region contains numerous young, massive stars and  clusters such as Trumpler 14 and Trumpler 16, which are embedded within dense molecular gas and dust, offering critical insights into the mechanisms of the mass star formation and the role of stellar feedback in the dispersal of natal molecular clouds \citep{Preibisch12}. A well-known tracer of recent star formation is the broad group of young stellar objects (YSOs). This class consists of two main groups of objects: protostars and pre-main sequence stars (PMS). 
One of the most prominent features of YSOs is their photometric variability, as well as some characteristic spectral signatures, like emission lines. Generally, the variability of the YSOs can be divided into two broad categories: caused by extrinsic activity due to an accretion disk or caused by magnetic activity producing significant spots on the photosphere. 

Several hundred studies have been conducted in the Carina region. We highlight some of the most relevant to our investigation of the YSOs of the region. \citet{Reipurth01} work provided a comprehensive overview of the Carina Nebula, emphasizing the impact of proto-stellar jets and outflows in shaping the morphology and dynamics of the nebular environment. \citet{Povich11} presented a catalog of 1439 YSOs in the Carina Nebula, identified via mid-infrared excess emission associated with dusty circumstellar disks and envelopes. The catalog includes Herbig Ae/Be stars, as well as their less evolved progenitors. \citet{Gaczkowski13} reported 642 Herschel-detected sources, predominantly classified as Class 0 protostars. \citet{Zeidler16} conducted a deep, wide-field near-infrared survey of the entire Carina Nebula Complex using the VISTA telescope, identifying 8781 sources with strong infrared excess, which were classified as candidate YSOs. More recently, \cite{2023JKAS...56...97H}  identified 3331 pre-main sequence (PMS) members and 14,974 PMS candidates down to a limiting magnitude of $V$=22mag, based on spectrophotometric properties across infrared, optical, and X-ray wavelengths. In the present work, we also used the YSO catalogs for the Carina region compiled by    \cite{2014A&A...567A.109K}, \cite{2018A&A...620A.172Z},  \cite{2019MNRAS.487.2522M}, \cite{2021AJ....162..153N}, \cite{2021ApJ...917...23K},
\cite{Kuhn21},
\cite{Marton23}, and  \cite{Campbell23}. 

Despite extensive observational and theoretical efforts, and significant progress in understanding stellar formation, several fundamental questions remain unsolved—largely due to observational limitations. For instance, no specific outburst-triggering mechanism has been definitively ruled out; all proposed scenarios remain viable. Rather than attempting to isolate a single dominant physical driver, the contemporary approach emphasizes the identification and classification of shared observational features. A major limitation has historically been the small size of available YSO samples. In recent years, this constraint has been mitigated by the emergence of high-quality, multiwavelength, time-domain  surveys, such as those of the Zwicky Transient
Facility \citep[ZTF;][]{2019PASP..131a8003M}, the VISTA Variables in the Vía Láctea (VVV) and VISTA Variables in
the Vía Láctea eXtended (VVVX) surveys  \citep{Minniti10,Saito12,Saito24}, NEOWISE \citep{2014ApJ...792...30M}, and Gaia \citep{2016A&A...595A...1G, 2023A&A...674A...1G} which collected extensive datasets. These large-scale surveys, for example, enable a statistical characterization of the mass accretion process, which can now be pursued using modern machine learning and deep learning techniques. By analyzing variability patterns in light curves, it is possible to differentiate between underlying mechanisms such as accretion, stellar spots, and extinction-related variability \citep{Cody17}.
On the other hand, spectroscopic follow-up on YSOs, particularly in the near-infrared regime, remains scarce. Existing efforts, such as the IN-SYNC and  YSO group projects within the Sloan Digital Sky Survey  \citep[SDSS;][]{Cottle18, Kounkel18, Santana21, Campbell23, Kounkel23} and the spectroscopic follow-up on some other spectrographs \citep[see e.g.,][]{Contreras17a,Contreras17b, Guo21, Guo24}, represent important advances, but mid- to high-resolution spectroscopy is still not very common. Such data are critical for validation of the proposed photometric classification and constraining the underlying physics.

In this study, we combine the variability analysis from the VISTA Variables in the VVVX \citep{Saito24} with the medium-resolution $H$-band spectroscopy from the APOGEE-2 survey \citep{Majewski17}, covering 766 stars. Our objective is to validate the classification of YSOs by eliminating contaminants such as binary stars, asymptotic giant branch (AGB) stars, novae, and long-period variables, which can exhibit similar light-curve characteristics, particularly among high-amplitude variables. We also search for spectral signatures indicative of high accretion rates, such as prominent emission features in the $H$-band spectra.

This first paper in our series presents a catalog of confirmed YSOs in the Carina region. We include analyses of their spatial distribution, effective temperatures, mass estimates, and photometric variability. A more detailed characterization of the sample, including outliers with unusual photometric and spectroscopic behavior, will be presented in a forthcoming follow-up paper.

\section{Observations and Data Reduction}\label{section2}
The target sample of the YSOs for spectroscopic observation was prepared on the basis of the near infrared variable star candidates. These were obtained from proceeding $K_S$-band images taken from  the VVVX ESO near-infrared public survey \citep[]{Saito24}, acquired up to 2019 June.  Subsequently, the variable candidates were cross-matched with existing YSO catalogs (up to 2020) from \citet{Povich11}, \citet{Preibisch11}, and \citet{Zeidler16}. This procedure defined two samples: a “variable” sample, containing objects exhibiting $K_S$-band variability exceeding 0.2mag in $K_S$, and a “non-variable” sample. The final catalog of 1600 objects were proposed for spectroscopic follow-up with the Sloan Digital Sky Survey IV (SDSS-IV) APOGEE-2 infrared spectrograph.

APOGEE-2 \citep{Majewski17} is a second-generation, multiobject near-infrared spectrograph mounted on the 2.5m du Pont Telescope at Las Campanas Observatory, Chile \citep{Gunn06}. As the successor to the original APOGEE instrument, it operates as part of the SDSS IV \citep{Blanton17}. APOGEE-2 covers only the spectral range of 1.51–1.70$\,\mu$m at a resolving power of R\,=\,22500. Each plate allows simultaneous observation of approximately 300 targets, with a typical fiber allocation of 250 for science targets, 35 for sky background, and 15 for telluric standard stars. The minimum separation between fibers is limited by a collision radius of 70$''$. APOGEE-2 observations of the Carina star-forming region were conducted on 2020 March 5, as part of an external CNTAC program (by principal investigator N. Medina). In total, 766 spectra were obtained. The data were processed under Data Release 17 \citep[DR17; see][for more details]{Abdurrouf22}. DR17 is the final data release of the fourth phase of the SDSS-IV and contains observations through 2021 January, including for the first time the southern infrared spectroscopy from  APOGEE-2. The preliminary reduction, the stellar parameters, and abundances are determined using the APOGEE Stellar Parameters and Chemical Abundance Pipeline \citep[ASPCAP; ][]{GarciaP16}, which relies on the FERRE optimization code \citep{AllendeP06}. 

In 2020, at the time of our APOGEE-2 observations, only a few epochs of VVVX were available. As mentioned previously, the VVVX is an ESO near-infrared public survey \citep{Saito24} performed on the 4\,m VISTA telescope at Cerro Paranal Observatory, Chile. This is an extension of the VVV survey \citep{Minniti10, Saito12}.
The VVV  mapped  562\,deg$^{2}$ in the Galactic bulge and the southern disk in five band near-infrared (NIR) broadband filters: $Z$, $Y$, $J$, $H$, and $K_S$, which has a time coverage spanning five years, between 2010 and 2015. The observations were performed with the VIRCAM NIR camera \citep{Dalton06}, with an array of 16 detectors with 2048\,$\times$\,2048 pixels each. The disk area is divided into 152 observing areas (1.5\,$\times$\,1.2 deg each, called {\it tiles}) and the bulge is covered with 196 tiles. The VVVX survey expanded the area of the original VVV footprint in both Galactic longitude and latitude, with an area of 480\,$\rm deg^2$ in the Galactic bulge plus 1170\,$\rm deg^2$ in the inner plane (including the original VVV), from $l$\,=\,$-$130\,deg to $l$\,=\,+20\,deg (7\,hr\,$<$\,R.A.\,$<$\,19\,hr). The VVVX survey started in 2016 and finished in 2023 \citep{Saito24}. The multi epoch observations were performed only in $K_S$ and up to 46 epochs are available. As the VVV survey, the VVVX produces unevenly spaced light curves, which provides some challenges, but in most of the cases, the variability can be detected and studied. 

According to wide-field sub-mm survey of \cite{Preibisch11}, the size of the Carina Nebula Complex is about 50\,pc (at 2.6\,kpc), corresponding to an extent of 1.25\,{$\rm deg^2$ on the sky. The complete region is covered by 6 VVVX tiles, namely e1040, e1041, e1085, e1086, e1130, and e1131, with a total area of $\sim9$\,{\rm deg$^{2}$}. For illustration, the selected regions and the total field of view are shown in Figure~\ref{FoV}.

\begin{figure*}[htbp]
\begin{center}
\includegraphics[height=10.5 cm,width=10.5 cm]{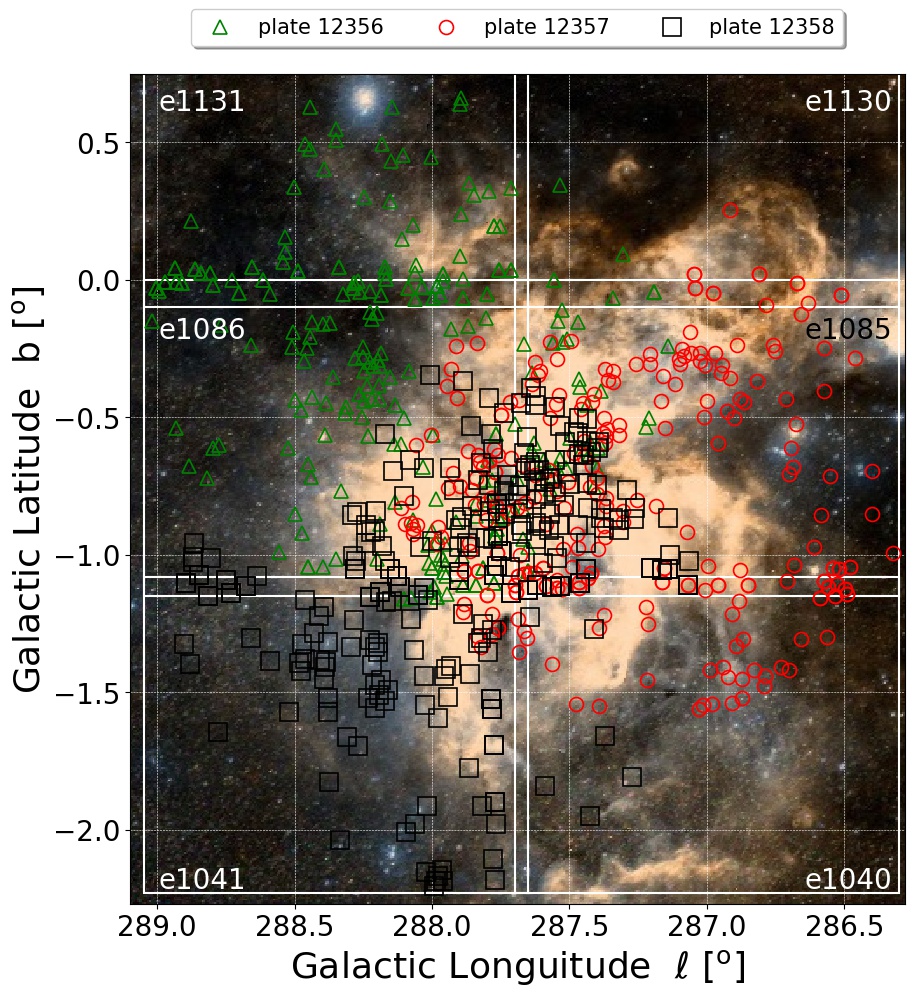}
\caption{The Carina star-forming complex region. The labeled six observed tiles: e1040, e1041, e1085, e1086, e1130 and e1131 from the VVVX survey. The symbols indicate targets observed on each of the three APOGEE-2 observed plates. In the background, SDSS2 color image. Galactic north is up, galactic east is to the left.}
\label{FoV}
\end{center}
\end{figure*}

The  variable stars are detected using the automated tool from \citet{Medina18}. 
Briefly, each pawprint image was retrieved from the Cambridge Astronomical Survey Unit\footnote{http://casu.ast.cam.ac.uk/} (CASU). Then point spread function (PSF) photometry was obtained using the $\mathtt{Dophot}$ software ~\citep{2012AJ....143...70A} 
in all available images. The calibration process for the VISTA system was done using the aperture photometry catalogs produced by the CASU \citep[for more details, see][]{Medina18, Borissova19}. The 6 VVVX fields produced 6,352,116 stars with PSF photometry.    
Two main variability indices are selected: the total amplitude {$ \Delta K_{\rm S}$ \citep{Contreras17a, Medina18} and the $\eta$} index \citep{vonNeumann41, Sokolovsky17, Medina18}, because they captured two fundamental properties of variable sources: the maximum change in brightness and the level of correlation among consecutive observations. The ranges of values for each index are determined from their dispersion. The $\Delta K_{\rm S}$ distribution was characterized using a nonparametric fit to determine the behavior of $\Delta K_{\rm S}$ as a function of $\overline{K}_{\rm S}$. Then the dispersion, $\sigma$, in $\Delta K_{\rm S}$ as a function of $\overline{K}_{\rm S}$ was measured. Sources with amplitudes above 4$\sigma$ were selected. For the $\eta$ index, we assumed that the index comes from a Gaussian distribution and thus used the $\sigma$ parameter of the fitted distribution as a proxy for the standard deviation. Sources more than 3$\sigma$ below the mean were considered, given the fact that an $\eta$ value that tends toward zero is a strong indicator of variability. 
More specifically, in this study, we selected any star with $\Delta K_{\rm S}>0.1$\,mag  AND $\eta$ values $<0.95$ as a variable source.
As a result, we identified 606 variable stars in the $K_S$ band, based on between 4 and 46 epochs.

\section{Classification of the APOGEE-2 $H$-band spectra}

To classify the APOGEE-2 spectra, we focused on the wavelength interval between 16795 and 16830\,\r{A}, which includes the characteristic Bracket\,11 (Br\,11) line and surrounding continuum. This line is a well-known tracer of circumstellar activity in YSOs \citep{Campbell23}. Some typical examples of Br\,11  profiles in our sample are shown in Figure\,\ref{Br11_examples}.

\begin{figure*}[htbp]
\begin{center}
\includegraphics[height=8.5cm,width=13cm]{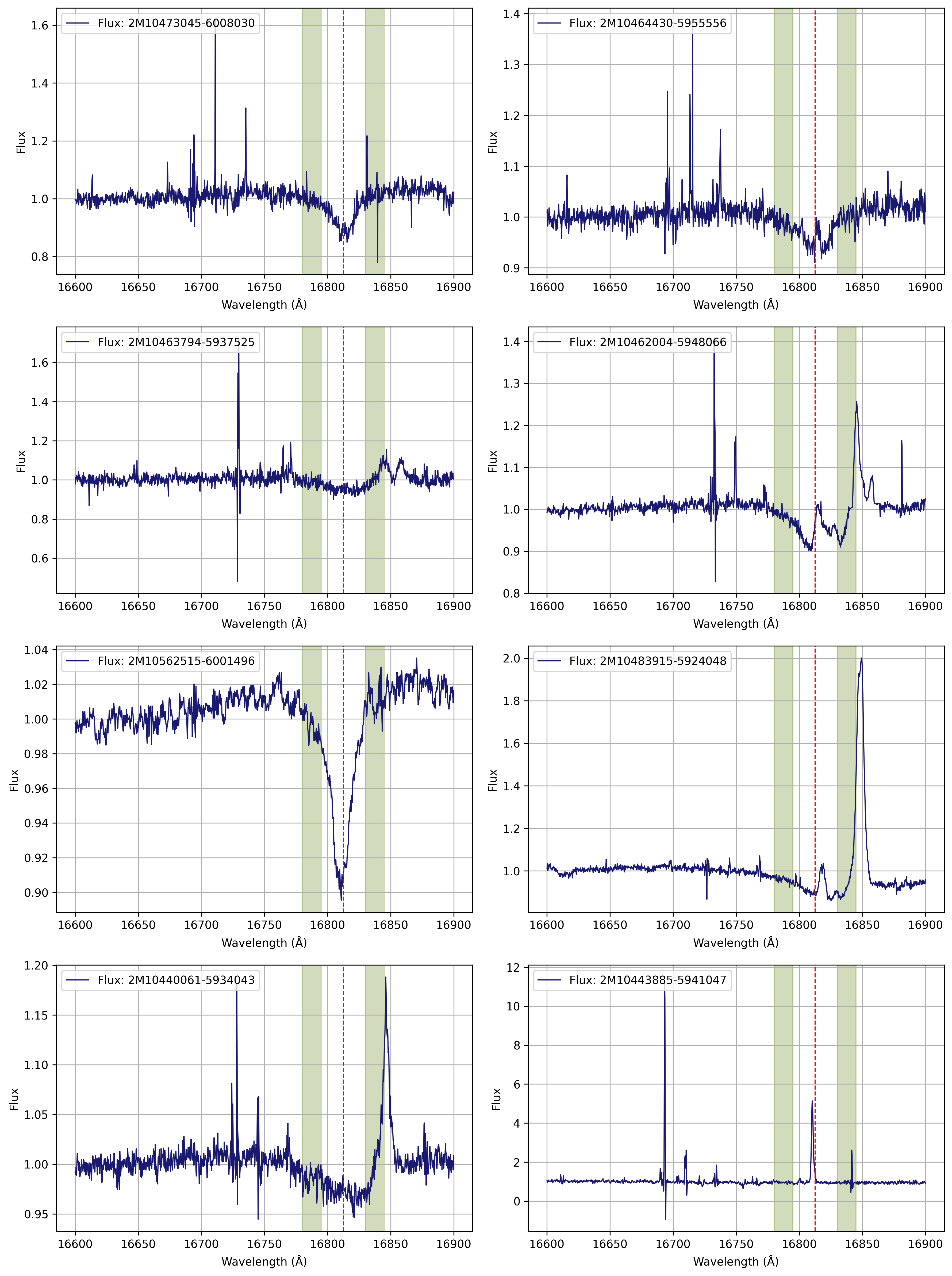}
 \caption{Typical APOGEE 2 $H$-band spectra around Br\,11 region. The spectra are normalized to the continuum by ASPCAP. The red line shows the  wavelength of Br\,11 line in the air. The green lines show 16795 to 16830 \r{A} wavelengths interval.}
\label{Br11_examples}
\end{center}
\end{figure*}

We applied the K-means clustering algorithm based on Br\,11 line shapes, using wavelength and flux values extracted from ASPCAP FITS files \citep{GarciaP16}. 
This simple, unsupervised method is chosen because it can effectively separate the data into clusters with minimal internal variance. 
The spectral data served as an input to the K-means clustering algorithm and are structured as a matrix composed of 766 rows, each representing an individual spectrum, and 151 columns, corresponding to flux values at specific wavelengths within the previously specified interval. Each column directly represents the flux at a given wavelength step, extracted from the normalized spectra. Consequently, each spectrum is characterized as a 151-dimensional vector in parameter space.
To mitigate biases arising from magnitude differences, flux values are standardized using the StandardScaler() class from Python’s Scikit-learn library.
This is a critical preprocessing
step to ensure that each variable (i.e., the flux of each wavelength) possesses a mean of zero and a standard deviation of one, thus normalizing the scale of the input features and preventing variables with larger magnitudes from disproportionately influencing the clustering process. 
The model was tested with various numbers of clusters, and we selected eight as the optimal choice. The other key hyperparameters like ninit=10, maxiter=1000, and init\,=\,'k-means++' are configured during the process, to the point that no significant improvements were observed by modifying them. From the eight clusters, three concentrated 98 $\%$ of the stars, corresponding to Br\,11 Emission, Absorption, and Unclassified/Noise groups. The algorithm stabilized after 24 runs, likely reaching a local minimum, but still provided a useful initial classification.

Since we expected that the full shape of the Br\,11 line could be discriminative, we decided to retain all 151 wavelength points as features to preserve detailed line structure. The dimensionality reduction (e.g., t-SNE) is shown on Figure\,\ref{k_mean} for visualization purposes. The 2D t-SNE projection of the stellar spectra, colored according to their K-means cluster assignment, plots the spatial distribution of the data. As can be seen, Clusters 1, 4, and 6 form well-defined and compact groups, indicating that the K-means algorithm has identified coherent structures in the data. Clusters 0 and 7 are more dispersed but remain coherent, while Clusters 2, 3, and 5 (each containing a single object) are likely outliers or atypical cases (for example, the Cluster 5 reflects a data error due to zero flux).
The right panel of Figure\,\ref{k_mean} shows the 2D t-SNE projection of the stellar spectra, colored according to the maximum flux value in the Br\,11 spectral line. A clear gradient in intensity is observed across the projection, with Cluster 1 containing the majority of high-flux sources, suggesting a population with strong Br\,11 emission. In contrast, Clusters 4 and 6 are dominated by lower flux values indicating absorption or more quiescent spectral profiles. This color mapping highlights the physical relevance of the clustering, as the algorithm has grouped spectra not only by shape but also by line flux  strength.
\begin{figure}[htbp]
\begin{center}
\includegraphics[height=7cm,width=\columnwidth]{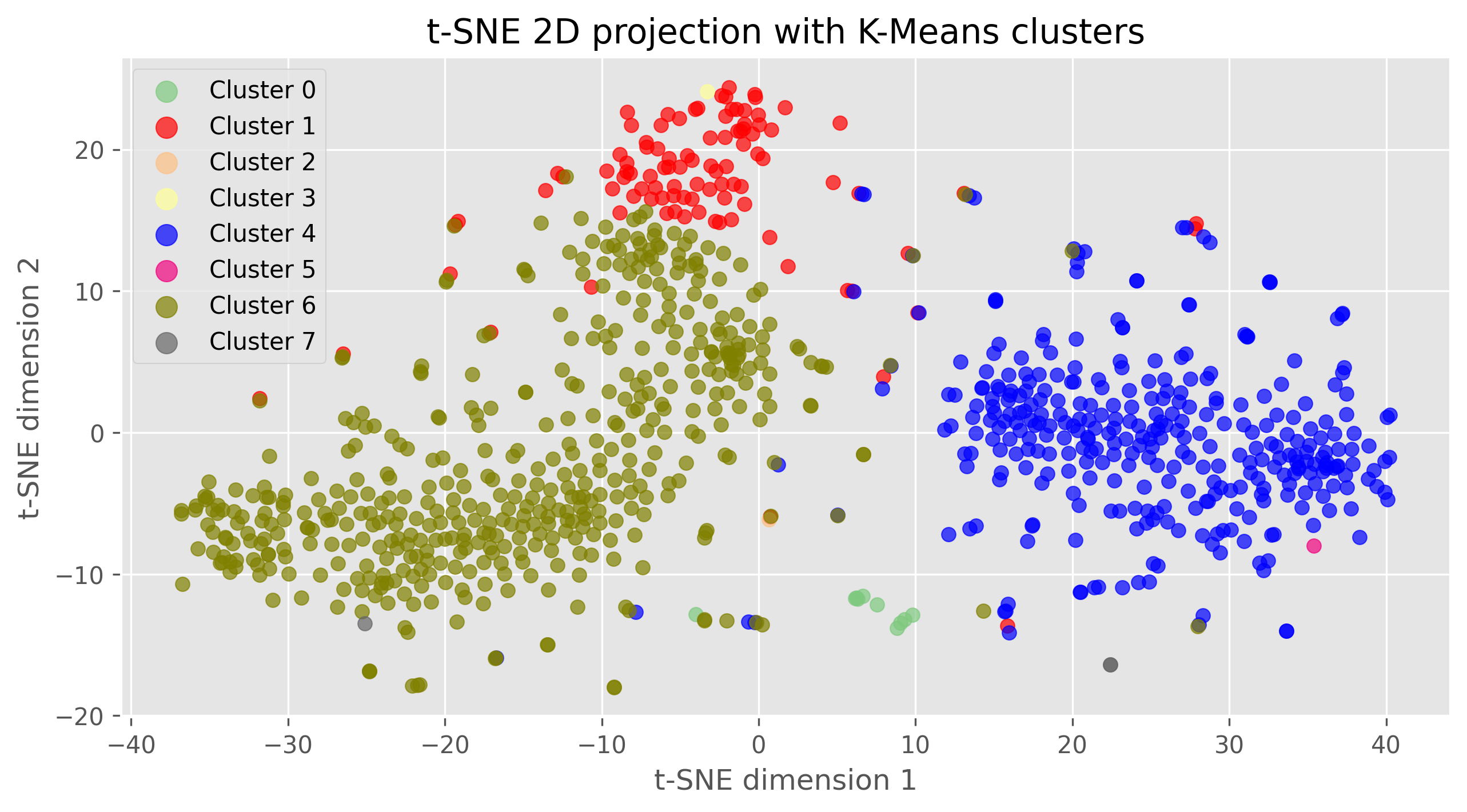} 
\includegraphics[height=7cm,width=\columnwidth]{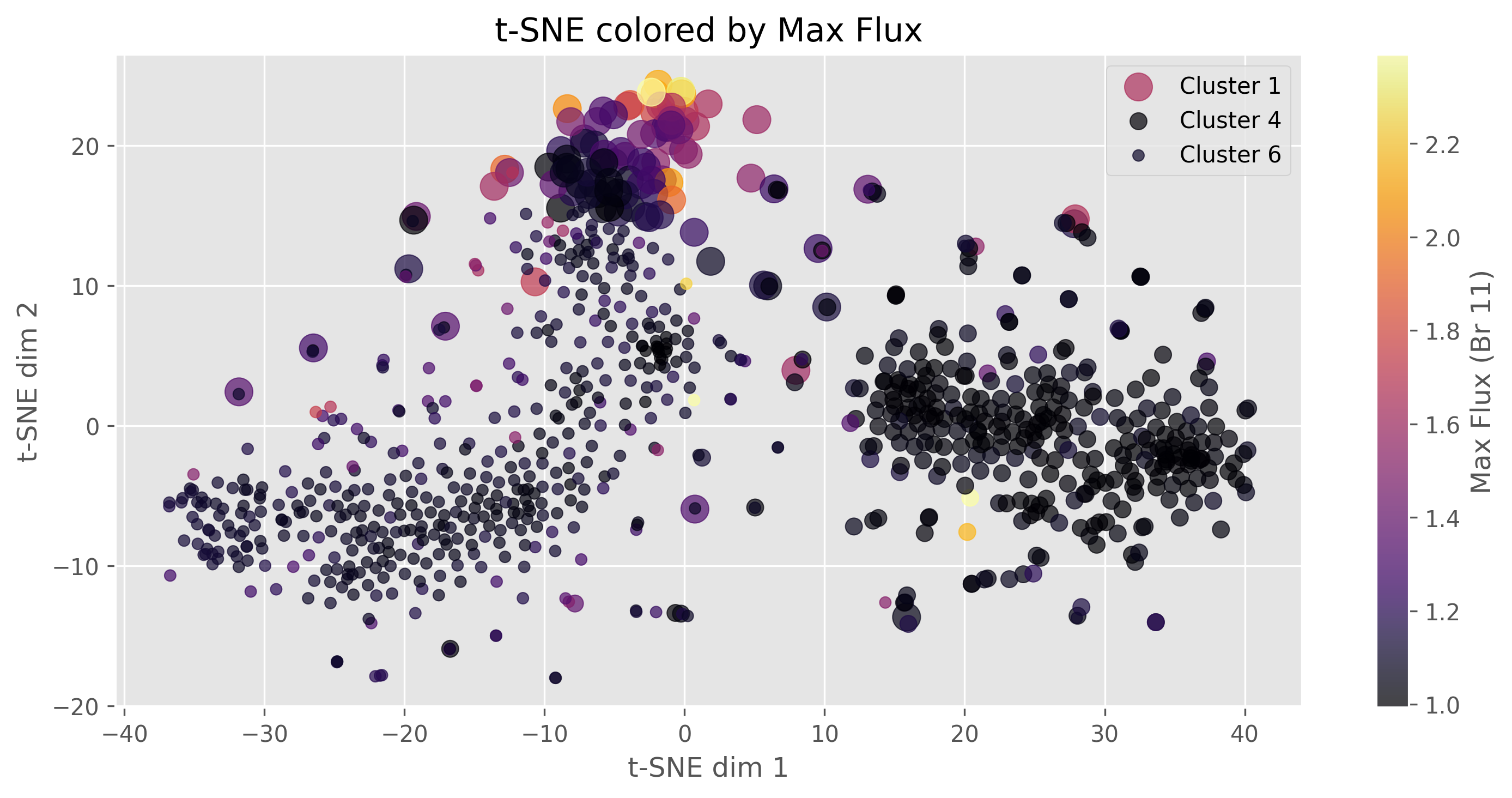} 
\caption{Left: The 2D t-SNE projection of the stellar spectra, colored according to their K-means cluster assignment. Right: The 2D t-SNE projection of the stellar spectra, colored according to the maximum flux value in the  Br\,11  spectral line. The colors and symbols are labeled.}
\end{center}
\label{k_mean}
\end{figure}

The physical differences between such identified clusters can be compared, calculating some summary statistics of key Br\,11 line parameters. Table\,\ref{K_mean_clusters} lists the median, standard Deviation (STD), and median absolute deviation (MAD) for the integrated flux for each group. 
Clusters 1, 4, and 6 show distinct median values ranging from 143 to 153, confirming that the K-means algorithm separated physically different line profile shapes, with Cluster\,1 showing the highest median integrated flux. The dispersion (STD and MAD) within these clusters is moderate, supporting the internal consistency of each group. Thus the statistics confirm that the structure of the clusters does not appear to be an artifact of the algorithm but rather reflects real differences in the  Br\,11 line (see Table\,\ref{K_mean_clusters} and Table\,\ref{high_mass}).

\begin{table*} \small
\caption{Statistics of the original K-means cluster groups and manual Reclassification Notes from the APOGEE-2 spectra.}             
\label{K_mean_clusters}      
\centering                          
\begin{tabular}{l c c c c l}        
\hline\hline                 
Name &      Number of &  Median         &  STD          &        MAD               & Notes \\
     &      K-mean Objects   &  Integrated Flux& Integrated Flux&   Integrated Flux&        \\
\hline    
Cluster 0	&	9	&	142.14&4.27&3.50& 2 moved to em,  3 moved to abs, 4 low S/N\\
Cluster 1	&	144	&153.19&2.94&1.64&	Br\,11 emission	\\
Cluster 2	&	1	&	164.61&0.0&0.0&low S/N,  removed	\\
Cluster 3	&	1	&	177.41&0.0&0.0&1 moved to em	\\
Cluster 4	&	224	&143.30&2.86&2.13&	Br\,11 absorption	\\
Cluster 5	&	1	&0.0&0.0&0.0&	no spectrum, removed	\\
Cluster 6	&	384&148.83&2.15&1.21&	unclassified	\\
Cluster 7	&	2	&145.34&1.21&0.56&	1 moved to abs, 1 EB star removed	\\
\hline   
\end{tabular}
\end{table*}

Although the  evaluation tools described herein are useful, they do not always reflect success for specific problems. In these cases, manual validation and domain knowledge are essential for adequately interpreting the results.  For manual validation, every individual spectrum in these eight clusters is visually inspected and subsequently  reclassified (if necessary) into three distinct clusters: Cluster\,1, containing stars exhibiting Br\,11 with emission; Cluster\,2 for stars with Br\,11 in absorption; and Cluster\,3, a heterogeneous group characterized by spectral classification uncertainties. The Cluster\,3 group uncertain classification can be due to a variety of spectral complexities, including low S/N, metal enrichment, radial velocity variations, P Cygni profiles, and the absence of Br\,11 in late-type stars. The reclassification notes are listed in the last column of Table\,\ref{K_mean_clusters}.
Thus after visual inspection we can calculate the classification accuracy of K-means algorithm. The confusion matrix diagonal elements show classification accuracy of 68\%, 69\%, and 89\% for the Cluster\,1, Cluster\,2, and Cluster\,3, respectively.

For the next step in our validation of the K-means classification of Cluster\,1 (stars with emission in the Br\,11 line), we compared  these objects with the \citet{Campbell23} catalog of the line measurements of 4255 Brackett emission-line sources from the APOGEE DR17 data set. Their study includes our fields of observation, namely the APOGEE plate numbers 12356, 12357, and 12358. The equivalent width (EW) values reported by \citet{Campbell23} for the Br lines were measured by a neural network constructed using tensor flow \citep{abadi2016}. The algorithm provided the probability thet (1) an emission line is present in the spectrum, (2) an emission line is present but appears to be double-peaked, and (3) there is no emission line in the spectral window. Since the goal of their work is to search for PMS accretors, the weak (EW of Br\,11 line $<$\,0.75\,\r{A}), Be, or nebular-like lines are excluded. The total number of emission objects selected by their algorithm  for our fields is 62. Comparison with Cluster 1 shows that we have 48 stars in common. The K-means algorithm incorrectly classified seven stars in the group of absorption  Br\,11 line stars (Cluster 2) and three in the unclassified group (Cluster 6). All of these show large radial velocities. The observed classification uncertainties suggest a limitation in the K-means approach, which can be improved by incorporating the radial velocity as an additional clustering parameter.
Thus Cluster\,1 has been adjusted to account for these corrections, and the final catalog of stars with Br\,11 line in emission contains 156 stars. In contrast with \cite{Campbell23} we consider all stars with emission lines (including double-peaked, P Cyg, likely Be stars and low-intensity ones).

As we have been pointed out,  Cluster\,3 is a heterogeneous group, which needs to be clarified using some additional information, not connected to our spectra. Thus a subsequent comparison was performed with previously published papers and their respective catalogs of PMS  stars and YSOs in the Carina region, described as follows: 

 \begin{enumerate}
     \item 
\citet{2014A&A...567A.109K} (hereafter K14) investigated the PMS stars in the Carina west region, combining optical, infrared, X-ray photometry and spectra. We have 33 objects in common, all classified by the authors as YSOs with ages $<$\,4.3 Myr, and low masses ($<$\,4.8\,\(M_\odot\)).

 \item 
A comparison with the \citet[][hereafter N21]{2021AJ....162..153N} catalog revealed 20 objects in common. The authors used X-ray and infrared observations to study the properties of three classes of young stars in the Carina Nebula: intermediate-mass (2–5 \(M_\odot\)) PMS stars (IMPS; i.e., intermediate-mass T Tauri stars), late-B and A stars on the zero-age main sequence (AB), and lower-mass T Tauri stars (TTS). All stars in common are classified by \citet{2021AJ....162..153N} as YSOs. Nine of them are IMPS,  10 are TTS, and 1 is unclassified. Only two stars are labeled as variables (2MASS J10421942-5950513 and 2MASS J10434698-5933182) by them, which we confirm with our $K_S$-band light curves. Another four stars, 2MASS J10440683-5936116, 2MASS J10454634-6000206, 2MASS J10461112-5952197, and 2MASS J10453834-5942078, are labeled as possible variables, which (with the exception of the last one) one we confirm.

 \item 
A comparison with the YSO catalogs of  \citet{2019MNRAS.487.2522M}  and \citet{Marton23} \citep[hereafter M19 and M23; see also][] {2016MNRAS.458.3479M}, resulted in 271 objects in common. Of these, 161 have spectra that the K-means classification places within Cluster\,3.

 \item 
A comparison with the X-ray-based catalog SPICY (\citep[][ hereafter Kuhn21]{Kuhn21} revealed 51 objects in common with our Cluster 3.

 \item 
We do not have any objects in common with the catalogs of \citet{2021ApJ...917...23K} and \citet{2021AJ....162..282M}, and have only two common stars (2MASS J10384536-5915447 and 2MASS J10505861-5957263) with \cite{2018A&A...620A.172Z}, Z18. 
  \item 
In the ASAS-SN Variable Stars Database \citep{2023MNRAS.519.5271C}, we found 12 common stars.
In addition, 23 more objects from the Cluster\,3 were found to be cataloged as YSOs in the SIMBAD astronomical database.
Some of our objects are found in different literature catalogs; in this case, we cited all of them.

\end{enumerate}

As a result of our findings, we have reclassified the 704 spectra in our sample with acceptable signal/noise and reliable spectral features into four groups: stars with emission in Br\,11 (labeled Emission-line YSOs); stars with absorption in Br\,11 (labeled Absorption-line YSOs); stars unclassified by the K-means algorithm (mainly spectra with a very weak (or no) Br\,11 line in absorption but many metallicity lines, labeled Unclassified); and stars without (or very weak) Br\,11, but classified as YSOs from the catalogs mentioned previously (labeled Literature). The remaining 62 stars with APOGEE-2 spectra are omitted due to their low signal/noise or unreliable spectral features.

 

The last step of our classification efforts was to include the $K_S$-band variability, taking into account that the variability is one of the most prominent features of the YSOs. As have been pointed out, we identified 606 variable stars in the $K_S$ band with amplitude range between 0.2 and 2.3 $K_S$ mag. Of these variables, 507 have APOGEE-2 spectra. 
Examples of eight $K_S$-band light curves are shown in Figure\,\ref{var_em_sp} for illustration. These stars show emission lines in the Br\,11 region, as well as relatively high amplitudes (Amp\,$>$\,1\,mag in $K_S$). They are previously classified as YSOs in the catalogs of  \cite{Kuhn21}, \cite{2019MNRAS.487.2522M}, and \cite{2023MNRAS.519.5271C}, but here we are reporting for the first time their 
strong near-infrared variability.
 
\begin{figure*}
\begin{center}
\includegraphics[width=\textwidth]{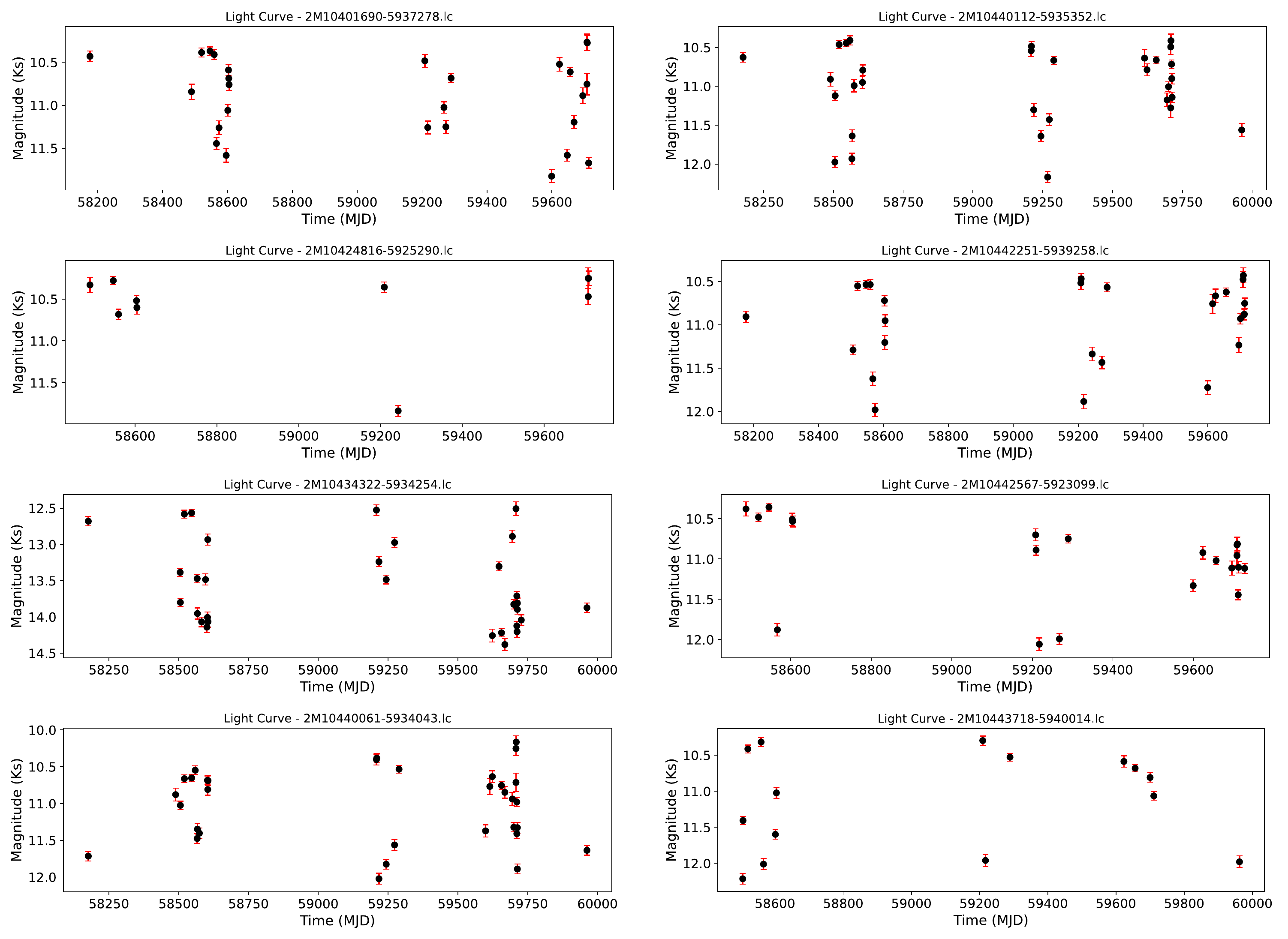}
\caption{Examples of the light curves of variable stars with $K_S$ amplitudes greater than 1.0 mag and emission in the Br\,11 line.}
\label{var_em_sp}
\end{center}
\end{figure*} 
 
In general, most of the Emission-line stars are variables (only 14\% do not show any variations in $K_S$). Around 67\% of the Absorption-line stars show some variability. 
Some 77\% of the Unclassified by K-means cluster group (144 stars) show variability in the $K_S$ band. We will consider all variable stars from the Unclassified cluster as confirmed YSOs. Subsequently, they are relabeled and added to the Literature group cluster with common label Literature/Variable.

We removed from our analysis the rest of the stars in Cluster\,3, which have uncertain spectral classification, are not found in literature studies of young stellar populations in Carina and do not show any variability in $K_S$.

We also used the catalog of double-lined spectroscopic binaries in DR17 APOGEE spectra by \citet{Kounkel21} and find five objects from our list to be identified as spectroscopic binaries in their work. These objects (2MASS J10454975-6019392, 2MASS J10444433-5945595, 2MASS J10463313-6009264, 2MASS J10503895-5922313, and 2MASS J10461677-6001345) are labeled in our final catalog.

Our final catalog is listed in Table\,\ref{catalog}, with  corresponding classification flags for Emissions (em), Absorptions (abs), Literature (lit), and Variable (var). 
While initial classification was performed using the purely unsupervised K-means algorithm, the subsequent visual inspection of the spectra, comparison with the literature catalogs, and addition of the photometric variability as an indicator of YSOs introduce a human-guided component. Thus we consider the final classification as a semisupervised method.


\section{Color--Magnitude Diagrams, Membership Probability, and Fundamental Parameters}

 The color-magnitude diagrams
 $K_S$  vs. $(J-K_S)$ from the VVVX and Gaia DR3 Gmag vs. (BP\,-\,RP) filters are plotted in Figure\,\ref{cmd_jk}. The PSF photometry in the near infrared was taken from Alonso-Garc\`ia  (2025, private communication). For the magnitude interval of $K_S$\,$<$\,10.5 mag (where most of the VVVX $K_S$ magnitudes are saturated), we added the 2MASS $J$, $H$, and $K_S$ magnitudes. These magnitudes and colors are not corrected for reddening. 
Each YSO is plotted with a colored symbol indicating which of
the three classification groups it has been assigned to.  In the $K_S$ vs. $(J-K_S)$ color magnitude diagram, the main sequence (MS) is represented by the relatively narrow vertical distribution of points between $0 \le (J-K_S) \le 0.8$. The Red Giant Branch (RGB) appears much broader at redder colors $(J-K_S) > 1$. The Emission-line YSOs are primarily concentrated around the redder regions of the CMD around $1.5 < (J-K_S) < 3$. Their distribution extends into regions where YSOs are typically found, coinciding with the spectroscopy and confirming these are active objects.  The Absorption-line YSOs are projected mainly on the MS region, confirming that these stars are more evolved and have lower levels of circumstellar material. The group of Literature/Variable YSOs are spread across the redder regions $(J-K_S) > 1.0$, indicating they might include objects with infrared excess.  

\begin{figure*}[htbp]
\begin{center}
\includegraphics[width=8cm]{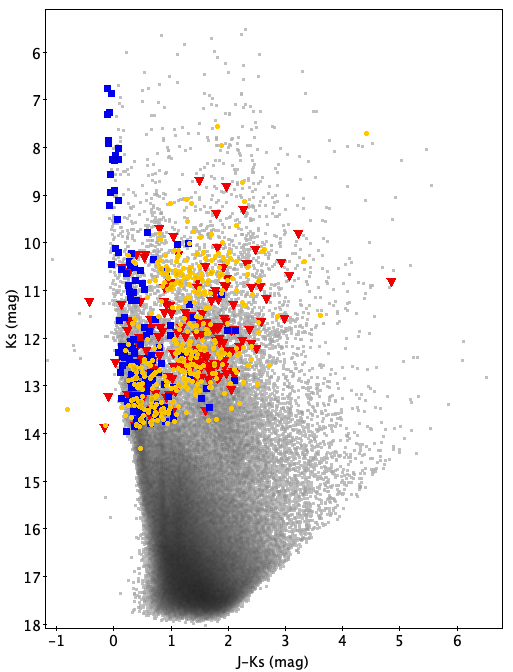}
\includegraphics[width=8cm]{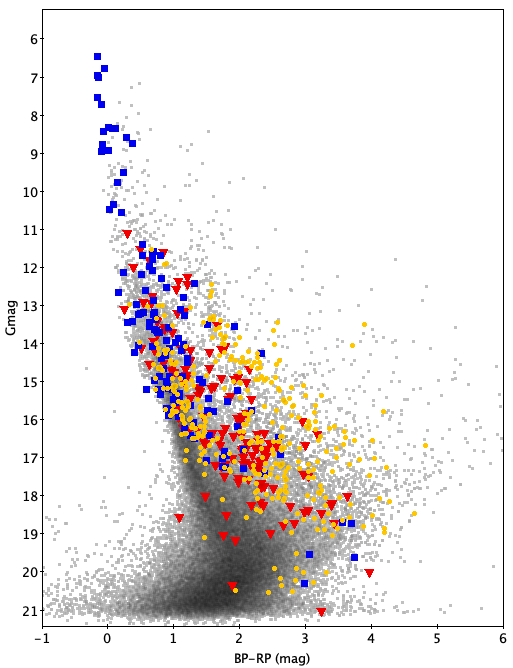}
\includegraphics[width=8cm]{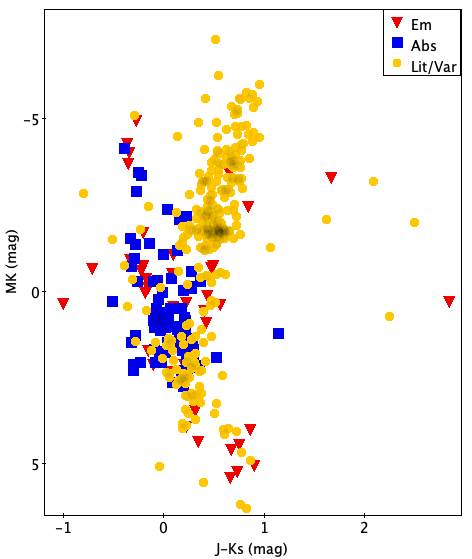}
\includegraphics[width=8cm]{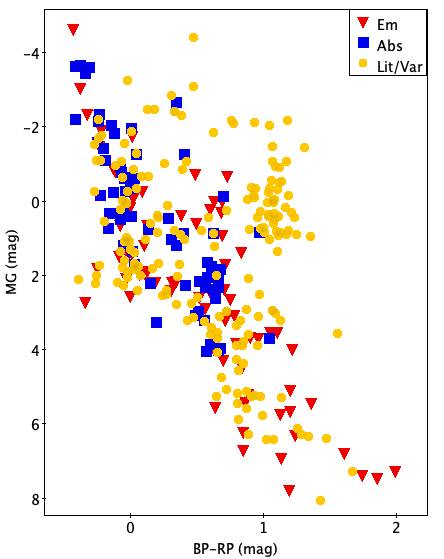}
 \caption{Upper panels: The $K_S$ vs. $(J-K_S)$;  G vs. (BP\,-\,RP)  color magnitude diagrams with YSOs of the Carina region overplotted. The magnitudes and colors are not corrected for the reddening and distance modulus. The color points are as follows: red triangles: Emission-line YSOs; blue squares: Absorption-line YSOs, yellow circles: Literature/Variable YSOs. Lower panels: Absolute magnitudes vs. corresponding dereddened colors in both systems. The colors and symbols are labeled. }
\label{cmd_jk}
\end{center}
\end{figure*} 

The distribution in the Gaia DR3 \citep{2023A&A...674A...1G}  $G$ vs. $(BP-RP)$ color-magnitude diagram confirms the trends derived from the near infrared diagram: Emission-Line YSOs and Literature/Variable YSOs dominate the reddest regions of the CMD. Their positions overlap with parts of the RGB and extend into the region where heavily reddened or embedded YSOs are expected. The Absorption-Line YSOs primarily occupy the MS and slightly extend into the RGB. Literature/Variable YSOs show a broader distribution, suggesting a population with diverse properties and indicating a mix of various stages of circumstellar interaction and more evolved YSOs.

In general, we confirmed  the classification from the K-means clustering and literature comparison for the young stellar populations in Carina.

\subsection{Membership probability}\label{membership}

To exclude possible nonmembers of the Carina star forming complex, we used the parallaxes, proper motion, and distances listed in the Gaia DR3 catalog \citep{2023A&A...674A...1G}, specifically the columns named: Plx,	ePlx, PM, pmRA, epmRA, pmDE, epmDE, and Dist.

\citet{refId1} calculated the distances of several selected young clusters in Carina using the parallaxes (in Gaia ERD3) of 237 spectroscopically identified OB stars. They found a mean distance of 2.36\,$\pm\,0.05$\,kpc for OB stars and 2.34\,$\pm\,0.06\,$kpc for a sample of X-ray-selected young stars. We applied their method to our sample of confirmed YSOs, listed in Table\,\ref{catalog}. 
For an initial estimate of the distances, we use the lineal fit of the parallaxes, as shown in  Figure\,\ref{pm_dis}. We do not apply any correction to the DR3 parallax data. The obtained value is 0.4451\,mas (2.25\,$\pm$\,0.11\,kpc). If we select only stars with errors of the parallaxes ($\sigma$) normalized by the median parallax of the sample ($\omega$) to be $< 0.3 $, as suggested by \citet{refId1}, then the median parallax value is calculated as 0.4228$\pm0.12$,  in very good agreement with their median value of 0.4242\,mas for OB stars. We will consider 59 stars with parallaxes bigger than 3$\sigma$ errors from the fit (0.36\,mas) as outliers.  
 
The rest of the stars (525) are further refined by analyzing the distance distribution taken from Gaia DR3 catalog  
(Figure\,\ref{pm_dis}). The median value is calculated as 2.37\,$\pm$\,0.11\,kpc.
Again, the stars within 3$\sigma$ errors from the median value are  considered probable members. For these stars, we calculate the mean pmRA and pmDEC as $-$6.3879\,$\pm$\,1.536 and 2.6441\,$\pm$\,1.291, respectively.
Finally, we plot the pmRA and pmDEC of all stars in our sample and select  the stars around these centers and 3$\sigma$ standard deviations. Thus 415 stars (67\%) are determined as the most probable members of the Carina complex.

\begin{figure}[htbp]
\begin{center}
 \includegraphics[width=8cm,height=4.0cm]{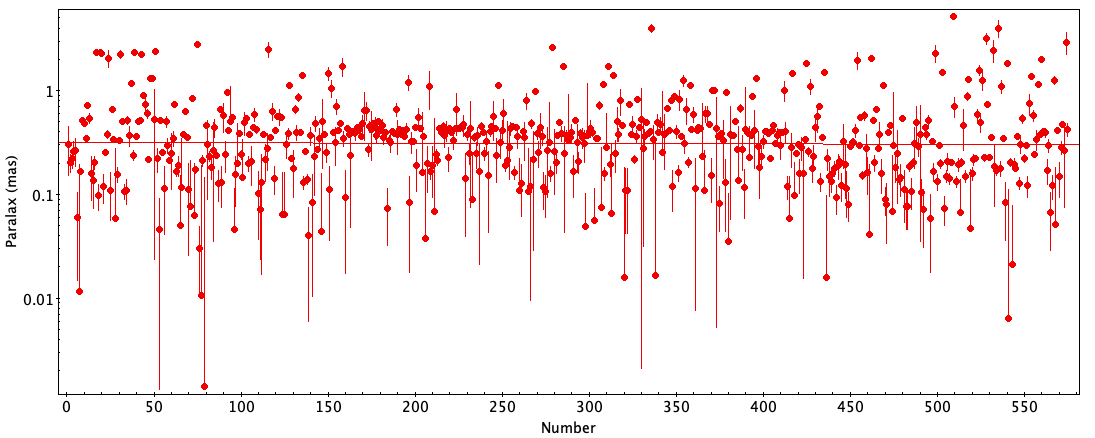}
\includegraphics[width=8 cm,height=4.0 cm]{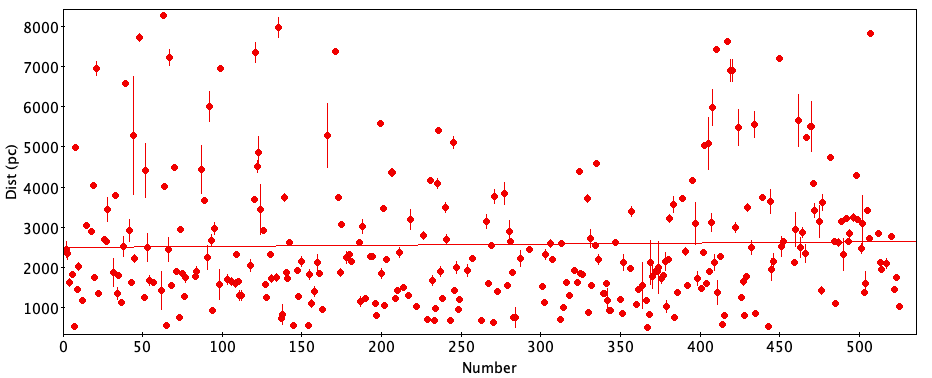}
\includegraphics[width=5.3 cm,height=4 cm]{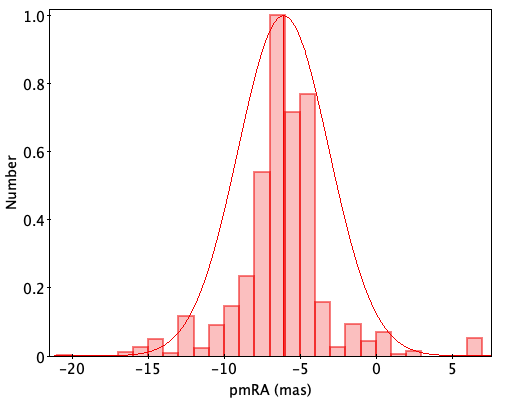}
\includegraphics[width=5.3 cm,height=4 cm]{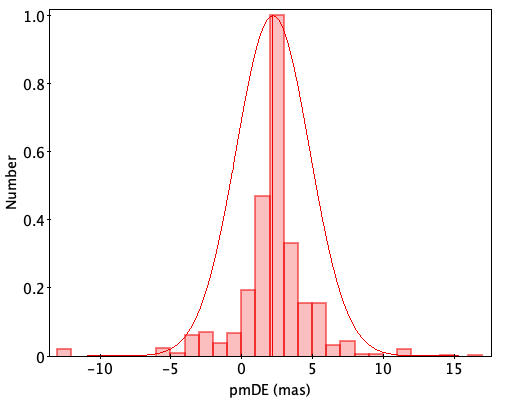}
\includegraphics[width=5.3 cm,height=4 cm]{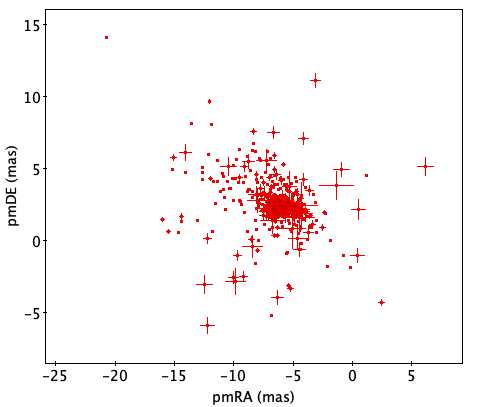}
 \caption{Parallaxes (in Log scale), distances and proper motions of the YSOs from Gaia DR3. The index numbers on the x-axis refer to the number of the star sorted by RA \citep{refId1}. The solid lines are the lineal and Gaussian fits plotted for for illustration. }
\label{pm_dis}
\end{center}
\end{figure} 

The estimated distances also help us discriminate the contamination from background stars, which is important because Carina is located right in the Galactic plane.  For example, we detect 45 stars with Distance $>$\,4000\,pc (approximately 7\% of the total sample), which are deemed to be background disk stars.

\subsection{Fundamental parameters}


The fundamental parameters of our stars (such as effective temperature, metallicity, stellar masses, etc.) can be obtained from their $H$-band spectra. There are  several value-added catalogs  available at \href{https://www.sdss.org/dr18/data_access/value-added-catalogs/}{https://www.sdss.org/dr18/data\_access/value-added-catalogs/}

 \begin{enumerate}
     \item 
The ASPCAP \citep{Abdurrouf22} standard output pipeline is not used because it is optimized mainly for cool stars.  
   \item 
APOGEE Net \citep{Sprague22}: APOGEE Net is a convolutional neural network, which has been calibrated to operate on all stars, including those for which the parameters have been challenging to derive by other means (such as PMS stars and OB stars). 
   \item 
APOGEE DR17 StarHorse: Distances, extinctions, and stellar parameters \citep{Queiroz20, Queiroz23}. This catalog combines high-resolution spectroscopic data from APOGEE DR17 with broadband photometric data from several sources (Pan-STARSS1, 2MASS, and AllWISE), as well as parallaxes from Gaia EDR3 \citep{refId0}, using the Bayesian isochrone-fitting code StarHorse  \citep{Queiroz18}. 
   \item 
The AstroNN: catalog of abundances, distances, and ages for APOGEE DR17 stars \citep{Leung19a,Leung19b}. This value-added catalog contains the results from applying the AstroNN deep learning code to APOGEE DR17 spectra to determine stellar parameters, individual stellar abundances (retrained with ASPCAP DR17) and distances (defined in Gaia EDR3).
 \end{enumerate}

To investigate the consistency of stellar parameters, we compared the effective temperatures and metallicities derived from three catalogs, APOGEE Net, StarHorse, and AstroNN, for our sample (Table\,\ref{catalog}). As can be seen from the box-plots in Figure\,\ref{Teff_comp} the median effective temperatures were found to be approximately 5000\,K, 6000\,K, and 7000\,K for APOGEE Net, StarHorse, and AstroNN, respectively. The AstroNN exhibited a systematic tendency toward higher temperature estimates and displayed the largest dispersion in temperature measurements. The presence of outliers exceeding 6000\,K suggests that none of the three catalogs is optimally calibrated for accurately determining the effective temperatures of young stars.
Regarding metallicity, the APOGEE Net catalog yielded a median value of approximately $-0.2$\,dex, while StarHorse and AstroNN produced higher median values, around $-0.1$\,dex and $-0.2$\,dex, respectively. All three metallicity distributions exhibited significant widths and the presence of outliers. Furthermore, the ages derived from AstroNN and Gaia DR3 catalogs are not reliable for our objects.  This is attributed to the fact that the training set for the respective neural networks mostly contains evolved stars.

\begin{figure}[htbp]
\begin{center}
\includegraphics[width=\columnwidth]{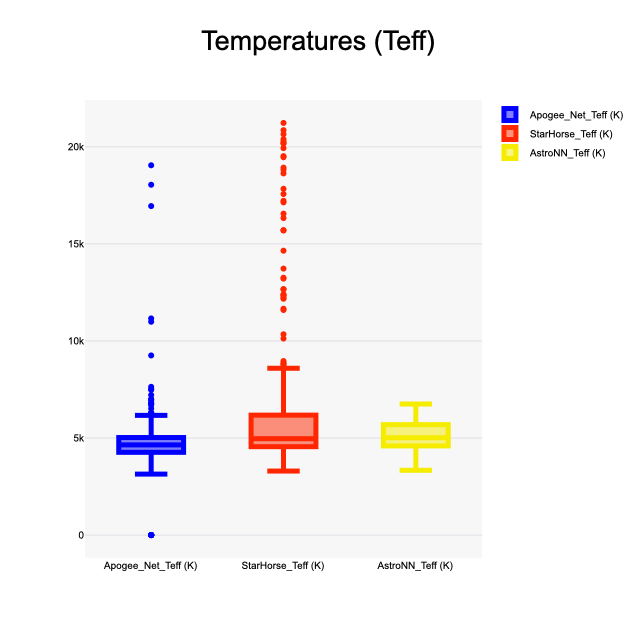}
\includegraphics[width=\columnwidth]{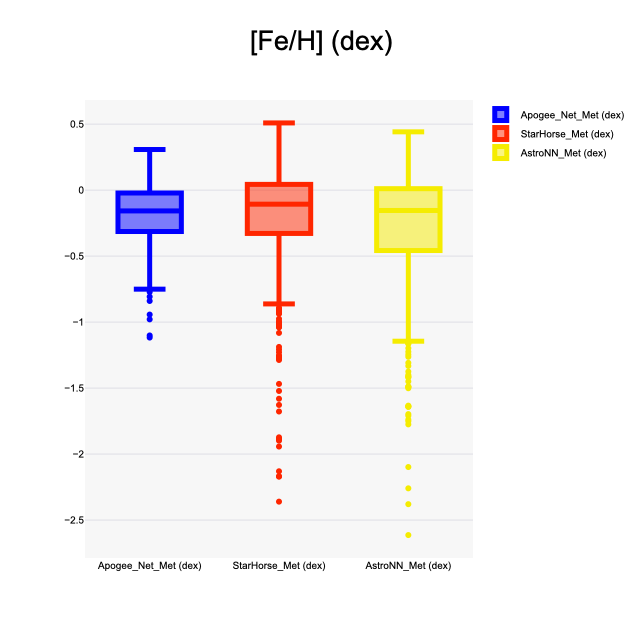}
\caption{Comparison of derived temperatures (in K) and metallicity from the value added catalogs APOGEE Net, StarHorse and AstroNN.}
\label{Teff_comp}
\end{center}
\end{figure}

Taking into account all these considerations, we selected the spectroscopically derived parameters provided by the  StarHorse value-added catalog for further analysis. The cross identification with our catalog (Table\,\ref{catalog}) produced 422 objects in common, and as suggested by \cite{Queiroz23}, we used the median value 50th percentile as the best estimate, with uncertainty determined using the 84th and 16th percentiles.

On the other hand, we can use the Gaia DR3  \citep{2023A&A...674A..39G}, provided physical parameters such as distance, reddening, temperature, metallicity, luminosity, and masses.

In the lower panels of Figure\,\ref{cmd_jk} we plotted the color-magnitude diagrams of the labeled samples of our objects (em, abs, lit/var) in the optical and near-infrared, corrected for individual reddening and distances, as derived by StarHorse and Gaia DR3, respectively.  We used the \cite{1989ApJ...345..245C} defined extinction coefficients. The distribution along the MS is clearly visible in both diagrams, as well as the intrinsically reddened stars.

 
 \subsection{Temperature, Metallicity, and Mass distributions}

Figure\,\ref{parameters} (upper panel) shows the StarHorse (left) and Gaia DR3 \citep{2023A&A...674A..39G} derived temperatures. As should be expected, the Emission- and Absorption-line YSOs are hotter (with temperatures higher then 10,000\,K) and more variable in temperature than  the Literature/Variable-identified YSOs.  The Gaia DR3 measured higher temperatures and showed more outliers. We do notice some clustering of the Emission-line objects, compared to the  Absorption-line and Literature/Variable objects, to the most active part of the Carina area. However, the Kolmogorov--Smirnov test does not show any statistically significant clustering among the different cluster groups.

\begin{figure*}[htbp]
\begin{center}
\hspace{0.1 cm}
 \includegraphics[width=\columnwidth]{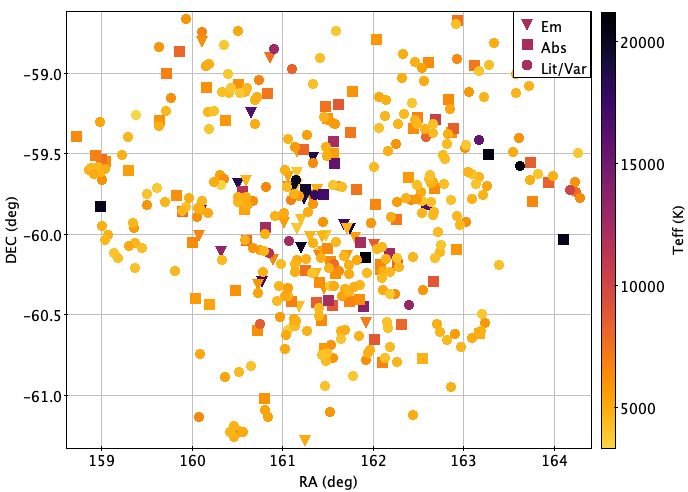}
\includegraphics[width=\columnwidth]{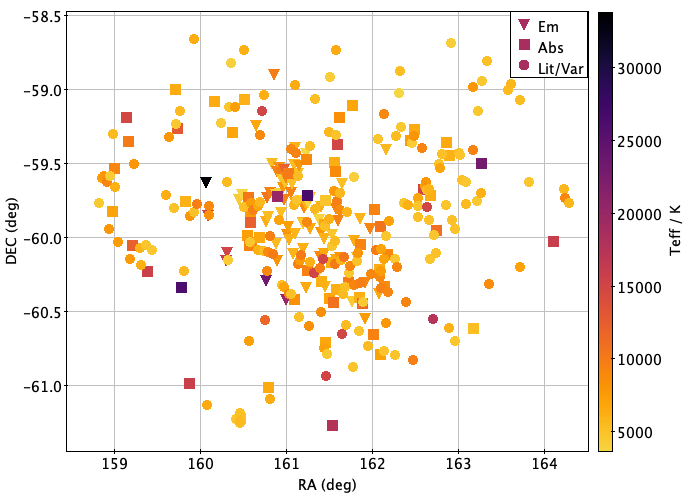}
 \includegraphics[width=\columnwidth]{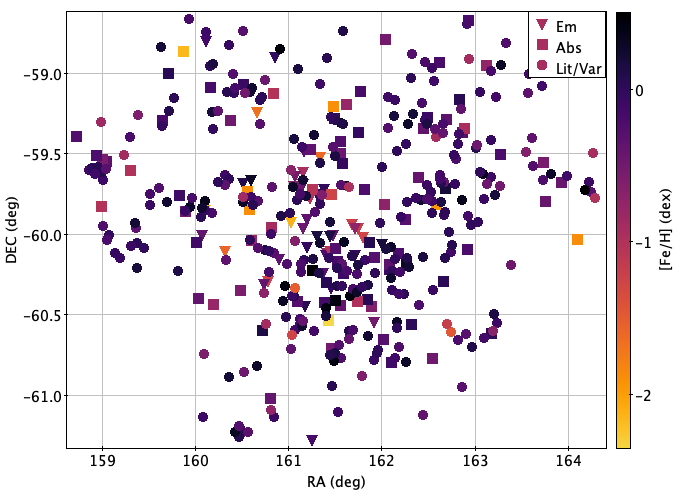}
 \includegraphics[width=\columnwidth]{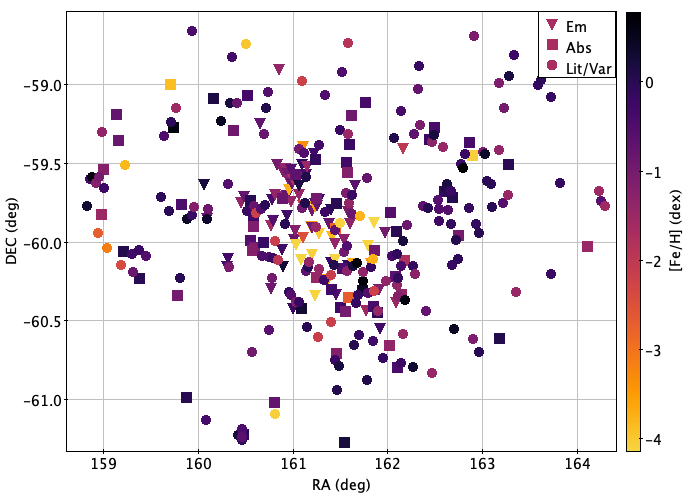}
  \includegraphics[width=\columnwidth]{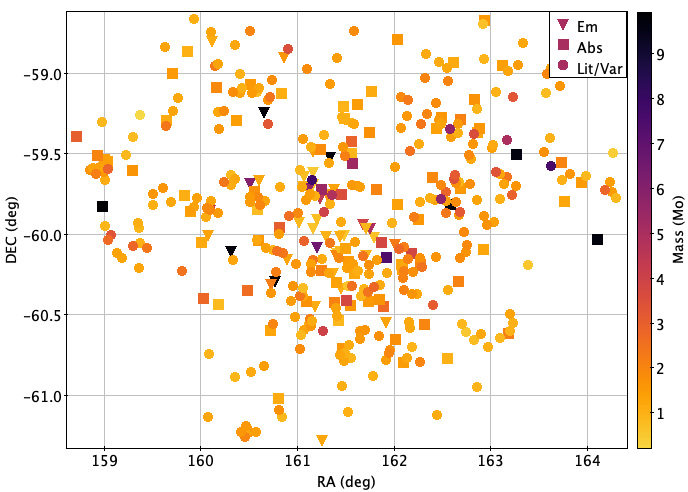}
\includegraphics[width=\columnwidth]{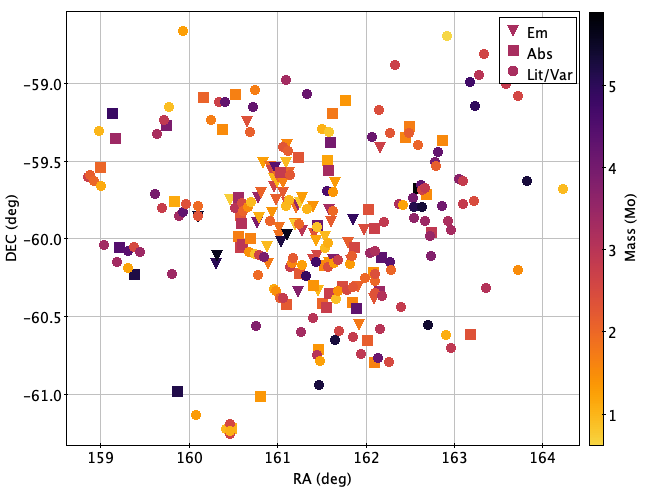}
 \caption{Upper panel: The StarHorse (left) and Gaia DR3 derived temperatures (right) of the YSOs. Middle panel: The StarHorse and Gaia DR3 derived [Fe/H] in dex. Lower panel: The StarHorse and Gaia DR3 derived masses. The colors and symbols are shown in the bar labels. }
\label{parameters}
\end{center}
\end{figure*}

The metallicities ([Fe/H], in dex) obtained from StarHorse and Gaia DR3 are shown in Figure\,\ref{parameters} (middle panel).
Most of the YSOs peak around solar metallicity (between $-0.2$ and $0.2$\,dex). Like other parameters (temperature, mass),  Gaia DR3 measured lower metallicities and more outliers.  This could be explained by the fact that most of our stars are faint (in the optical $G$, BP, and RP bands) and variable; thus the Gaia parameters are not well constrained in DR3. 
  
The mass distribution of our objects is shown in Figure\,\ref{parameters} (lower panel). Following \citet{Vioque23}, we divided the sample into three groups: high-mass stars (with the mass interval around 9-10\,\(M_\odot\), intermediate-mass stars (in the range 4–8\,\(M_\odot\), and low-mass stars (1–4,\(M_\odot\)).

As can be seen in Figure\,\ref{parameters} more than 90 \% of our objects are low-mass stars. 
The list of high-mass stars contains eight stars with masses around 9.8\,\(M_\odot\). Table\,\ref{high_mass} shows their StarHorse-derived parameters. 
 
All of them have high temperatures and low metallicity, and some of them are far from the derived to a Carina nebula distance of approx. 2.4 kpc. Four of them show Br\,11 in absorption; these are most likely early B stars. The 2MASS J10530463-5930198 also has high temperature, but the distance and higher metallicity suggested that the star is a probable member of the Carina complex. The 2MASS J10562515-6001496 is also probable member of Carina.
The stars 2MASS J10411710-6005589, 2MASS J10423820-5914138, 2MASS J10430638-6017130, and 2MASS J10452188-5931036 show all Brackett lines in strong emission and Br\,11 lines with P Cyg profile. Their distance values indicate that they are most probably projected in the field of view of Carina.  Nevertheless, we confirmed the classifications of \citep[][via  mid-infrared excess emission]{Povich11}  and \citep[][via Gaia EDR3]{Lebzelter23}  of these stars as YSOs, most probably Ae/Be stars.

The list of intermediate-mass stars contains 15 stars listed in Table\,\ref{high_mass}. All are confirmed YSOs, probably members of the Carina complex according to their estimated distances.

\begin{table*}\small\tabcolsep=2pt
\caption{StarHorse-derived Parameters for High and Intermediate-mass Stars}             
\label{high_mass}      
\centering                          
\begin{tabular}{lcclclcclc}        
\hline\hline                
OBJ ID &  R.A.& Decl.&Br\,11& 		Mass& $T$eff&log\,$g$&		[Fe/H] &	Dist &	$A_{V}$\\
\hline    
2M10355551-5949326	&	158.98133	&	-59.82575	&	abs	&	9.92$\pm$0.01	&	20282$\pm$36	&	3.33$\pm$0.01	&	-1.01$\pm$0.11	&	4.61$\pm$0.14	&	2.96$\pm$0.27	\\
2M10430638-6017130	&	160.77658	&	-60.28695	&	em	&	9.85$\pm$0.08	&	17219$\pm$60	&	3.01$\pm$0.01	&	-1.25$\pm$0.23	&	6.52$\pm$0.51	&	7.12$\pm$0.79	\\
2M10411710-6005589	&	160.32128	&	-60.09970	&	em	&	9.77$\pm$0.07	&	13257$\pm$98	&	2.54$\pm$0.02	&	-1.58$\pm$0.19	&	13.8$\pm$0.59	&	6.83$\pm$0.58	\\
2M10452188-5931036	&	161.34117	&	-59.51769	&	em	&	9.74$\pm$0.08	&	18812$\pm$95	&	3.17$\pm$0.01	&	-1.63$\pm$0.14	&	14.18$\pm$0.92	&	6.93$\pm$0.49	\\
2M10423820-5914138	&	160.65919	&	-59.23718	&	em	&	9.73$\pm$0.05	&	16330$\pm$86	&	2.9$\pm$0.01	&	-1.68$\pm$0.13	&	9.2$\pm$0.38	&	9.19$\pm$0.49	\\
2M10562515-6001496	&	164.10482	&	-60.03046	&	abs	&	9.72$\pm$0.05	&	20394$\pm$430	&	3.31$\pm$0.14	&	-1.87$\pm$0.26	&	1.88$\pm$0.35	&	1.34$\pm$0.31	\\
2M10502071-5948475	&	162.58632	&	-59.81320	&	abs	&	9.69$\pm$0.01	&	16557$\pm$44	&	2.92$\pm$0.01	&	-1.94$\pm$0.04	&	5.86$\pm$0.11	&	5.78$\pm$0.19	\\
2M10530463-5930198	&	163.26933	&	-59.50551	&	abs	&	9.64$\pm$0.19	&	20650$\pm$156	&	3.52$\pm$0.07	&	-0.21$\pm$0.15	&	1.51$\pm$0.13	&	1.58$\pm$0.19	\\
\hline 
2M10443718-5940014	&	161.15494	&	-59.66708	&	em	&	7.75$\pm$0.28	&	20178$\pm$262	&	3.90$\pm$0.06	&	0.23$\pm$0.07	&	2.27$\pm$0.19	&	2.56$\pm$0.62	\\
2M10543003-5934487	&	163.62516	&	-59.58021	&	abs	&	7.60$\pm$1.25	&	20857$\pm$1900	&	4.05$\pm$0.14	&	-0.26$\pm$0.28	&	2.64$\pm$0.46	&	11.60$\pm$0.82	\\
2M10474015-6008464	&	161.91731	&	-60.14624	&	abs	&	7.42$\pm$2.65	&	21230$\pm$3375	&	3.90$\pm$0.24	&	-0.28$\pm$0.36	&	4.09$\pm$1.66	&	6.07$\pm$0.56	\\
2M10444901-6004194	&	161.20423	&	-60.07208	&	em	&	6.63$\pm$0.57	&	19934$\pm$254	&	4.12$\pm$0.09	&	-0.04$\pm$0.15	&	4.96$\pm$0.76	&	4.78$\pm$0.59	\\
2M10420218-5940295	&	160.50912	&	-59.67487	&	em	&	6.04$\pm$0.11	&	18629$\pm$617	&	4.14$\pm$0.01	&	0.08$\pm$0.08	&	2.10$\pm$0.02	&	3.57$\pm$0.38	\\
2M10442897-5942343	&	161.12072	&	-59.70953	&	em	&	5.72$\pm$0.61	&	20177$\pm$569	&	4.28$\pm$0.12	&	-0.43$\pm$0.55	&	1.98$\pm$0.11	&	1.66$\pm$0.52	\\
2M10495852-5946576	&	162.49386	&	-59.78269	&	abs	&	5.67$\pm$0.09	&	5674$\pm$66	&	1.84$\pm$0.06	&	-0.20$\pm$0.24	&	4.86$\pm$0.43	&	9.77$\pm$0.33	\\
2M10501878-5920575	&	162.57825	&	-59.34932	&	abs	&	5.63$\pm$0.10	&	4642$\pm$0.10	&	1.56$\pm$	0.10	&	-0.34$\pm$0.10	&	2.99$\pm$0.31	&	5.90$\pm$0.16	\\
2M10445990-5943149	&	161.24962	&	-59.72081	&	abs	&	5.42$\pm$0.01	&	19522$\pm$87	&	3.86$\pm$0.01	&	-1.04$\pm$0.03	&	2.08$\pm$0.06	&	2.66$\pm$0.51	\\
2M10461754-5933348	&	161.57311	&	-59.55969	&	abs	&	5.29$\pm$0.14	&	15689$\pm$448	&	3.78$\pm$0.08	&	-0.05$\pm$0.15	&	5.47$\pm$0.56	&	5.09$\pm$0.35	\\
2M10524194-5924592	&	163.17476	&	-59.41646	&	abs	&	4.96$\pm$0.44	&	15707$\pm$225	&	3.93$\pm$0.16	&	-0.04$\pm$0.10	&	2.61$\pm$8.69	&	1.58$\pm$1.02	\\
2M10470063-5957242	&	161.75266	&	-59.95673	&	em	&	4.95$\pm$0.12	&	19462$\pm$483	&	4.10$\pm$0.06	&	-1.21$\pm$0.20	&	2.50$\pm$0.18	&	2.65$\pm$0.51	\\
2M10452586-5945368	&	161.35779	&	-59.76024	&	em	&	4.82$\pm$0.30	&	17836$\pm$260	&	4.08$\pm$0.12	&	-0.60$\pm$0.09	&	1.73$\pm$0.28	&	10.15$\pm$0.37	\\
2M10445682-5946106	&	161.23679	&	-59.76962	&	em	&	4.38$\pm$0.53	&	18928$\pm$1609	&	4.29$\pm$0.09	&	-1.04$\pm$0.21	&	2.06$\pm$0.11	&	2.27$\pm$1.02	\\
2M10402432-5950462	&	160.10134	&	-59.84619	&	em	&	4.10$\pm$0.03	&	14649$\pm$24	&	3.51$\pm$0.01	&	-2.17$\pm$0.25	&	4.17$\pm$0.09	&	5.23$\pm$0.20	\\
\hline   
\end{tabular}
\end{table*}

The Gaia DR3 determined the masses of a smaller number of stars, as well as lower values (up to 6\,\(M_\odot\)) compared to StarHorse (up to 10\,\(M_\odot\)). The mean mass of stars with emission in Br\,11 is around 2.4\,\(M_\odot\), and while the masses of the Absorption and Literature/Variable stars are similar, around 2.5-2.6\,\(M_\odot\), they show a larger spread compared to stars with  Br\,11 in emission.  The YSO masses are plotted throughout the R.A. and Decl. space, with no apparent strong clustering. Visually, the higher-mass objects (greater than 4 \(M_\odot\)) seem slightly more concentrated in the most active part of the region, while the lower-mass objects are more evenly spread throughout. However, the statistical Kolmogorov--Smirnov test does not show any clear correlation between mass and spatial position of the Carina's YSOs. 
 
In general, we do not find any statistically significant difference among the spatial distributions of the three groups (Emission, Absorption, and Literature/Variable YSOs) when separated according to their temperature, metallicity, and mass. 

\section {$K_S$ band variability indices} 

 We calculated the following statistical indices from the light curves of all 606 variable stars: mean\_mag, delta\_mag, amplitude (defined as the difference between the median value of the five highest and median value of the five lowest magnitudes), variance, std\_dev, skewness, kurtosis, slope\_index, gpv, structure\_function, time\_span, best\_period, and falce\_alarm\_probability (FAP), and also give the  num\_observations. The results are presented in Table\,\ref{catalog}. 
 The following is an assessment of the reliability of the statistical results obtained from the light curves: 
 \begin{itemize}
\item Number of observations $<$10: Statistical indices become quite unreliable. Results for these objects should be interpreted very cautiously as they are prone to biases and random noise.
\item Number of observations $10$-$25$: Basic statistics (mean, amplitude, variance, std deviation) are acceptable, but higher-order statistics (skewness, kurtosis, slope index, structure function) might still be marginally reliable, and should be interpreted with caution.
\item Number of observations $>$25: Most indices become significantly more reliable. Statistical confidence is better here, especially for structure functions and periodicity estimates.
\end{itemize}

\begin{table*}
\caption{Reliability of the Statistical Indices of the Objects with More Than 25 Epochs.}             
\label{stat_realibility}      
\centering                          
\begin{tabular}{l l l }        
\hline\hline                 
Index & Astrophysical Information & Reliability \\
\hline    
mean\_mag & Average brightness level & High \\
delta\_mag & Magnitude range (variability amplitude indicator) & Moderate \\
amplitude & Robust amplitude indicator (less sensitive to outliers than delta\_mag) & Moderate \\
variance / std\_dev & Degree of photometric variability & Moderate \\
skewness & Asymmetry in brightness distribution & Low/moderate \\
kurtosis & Indicator of outliers/extreme events & Low/moderate \\
slope\_index & Long-term monotonic brightness trend & Moderate \\
gpv & Gaussian Process Variability indicator (model-independent variability) & Moderate \\
structure\_function & Variability strength as function of time lag & Moderate \\
best\_period & Possible rotation/accretion period (stellar rotation, hot spots, accretion) & Moderate \\
\hline   
\end{tabular}
\end{table*}

As can be seen from Table\,\ref{catalog}, nearly half of the variable stars are foreground or background stars. In the following analysis, we focus our statistical study specifically on Carina. The Carina dataset comprises 329 variable objects, each with between 4 and 46 epochs of observation in the near-infrared $K_S$ filter. Only 6 objects have fewer than 10 epochs, while 31 objects have between 10 and 25 epochs. Some of the results are illustrated in Figure\,\ref{stats}.

The statistical analysis is influenced by two main factors: the number of observations (epochs) and the $K_S$ magnitudes. As was already mentioned, the brightest stars in the sample fall into the nonlinear response region of the detectors. The saturation level varies with the seeing conditions, but it is typically around $K_S \sim 10.5$. 
DOPHOT uses PSF fitting and can partially correct the flux of moderately saturated stars. Well suited to analysis of the heavily blended fields, it is unable to reliably fit the fluxes of heavily saturated sources and usually masks them instead. To ensure the robustness of our results, we introduced RELIABLE and UNRELIABLE flags. Stars are considered RELIABLE if they have $K_S$\,$\ge$\,11.5 and num\_observations $\ge$\,30. These 222 stars are marked in light green in Figure\,\ref{stats}. 

Period analysis was performed using the Lomb--Scargle method, which is well suited for irregularly sampled time-series data. For each source, we also calculated the FAP to assess the significance of the detected periods. Only stars with flag RELIABLE and FAP values below 0.2 were used for further analysis to ensure a minimum level of periodicity confidence. These 117 periodic variables, identified as probable Carina members, are marked with blue crosses in Figure\,\ref{stats}.

Histograms for the number of observations (epochs) per object are presented in the upper-left panel of Figure\,\ref{stats}. Red bars represent all sources, light green corresponds to reliable sources, and blue indicates periodic variables among the reliable ones. As shown, most variable stars have between 35 and 45 epochs.

The upper-right panel of Figure\,\ref{stats} presents an amplitude vs. mean $K_S$ magnitude diagram. This plot can help assess whether brighter or fainter stars exhibit higher variability amplitudes. Some stars display relatively large amplitudes, which may suggest stronger episodic accretion events, or more likely analyzing their individual light curves, rotational modulation caused by large stellar spots or structures within the circumstellar disk. Excluding the brightest UNRELIABLE stars (visible as the “plume” around $K_S \sim 11$), we observe a slight trend of increasing amplitude with decreasing brightness.

The skewness vs. kurtosis diagram, shown in the lower-left panel of Figure\,\ref{stats} highlights stars with non-Gaussian brightness distributions. Stars exhibiting both high skewness and high kurtosis are particularly interesting, as they may indicate episodic or eruptive variability.

Theoretically, kurtosis can take values from $-2$ to $\infty$. A kurtosis value near zero suggests a brightness distribution close to Gaussian (normal). Positive kurtosis ($>$0) reflects a more "peaked" distribution with heavier tails, implying a higher frequency of outliers or extreme brightness events --- potentially linked to episodic accretion or obscuration events in our sample. Conversely, negative kurtosis ($<$0) denotes a flatter distribution with lighter tails, indicating variability more symmetrically and evenly spread around the mean brightness. Such cases typically lack strong outliers and are relatively rare among strongly variable YSOs. In our dataset, these lower-kurtosis values are mostly associated with low-amplitude variables.

The final panel displays the kurtosis vs. $\log$(best period) relation. Although no clear trend is evident between kurtosis and the logarithm of the period, some stars with shorter periods do exhibit elevated kurtosis values. This may hint at short-period stars undergoing burst-like or episodic accretion or obscuration events. Many of these stars are noted as probable eclipsing stars or stars with fading events during the light curve visual check. This diagram serves as a useful tool for identifying periodic stars with burst-like or irregular behavior --- candidates that may warrant more detailed analysis or targeted follow-up observations.

\begin{figure*}
\begin{center}
 \includegraphics[width=\textwidth]{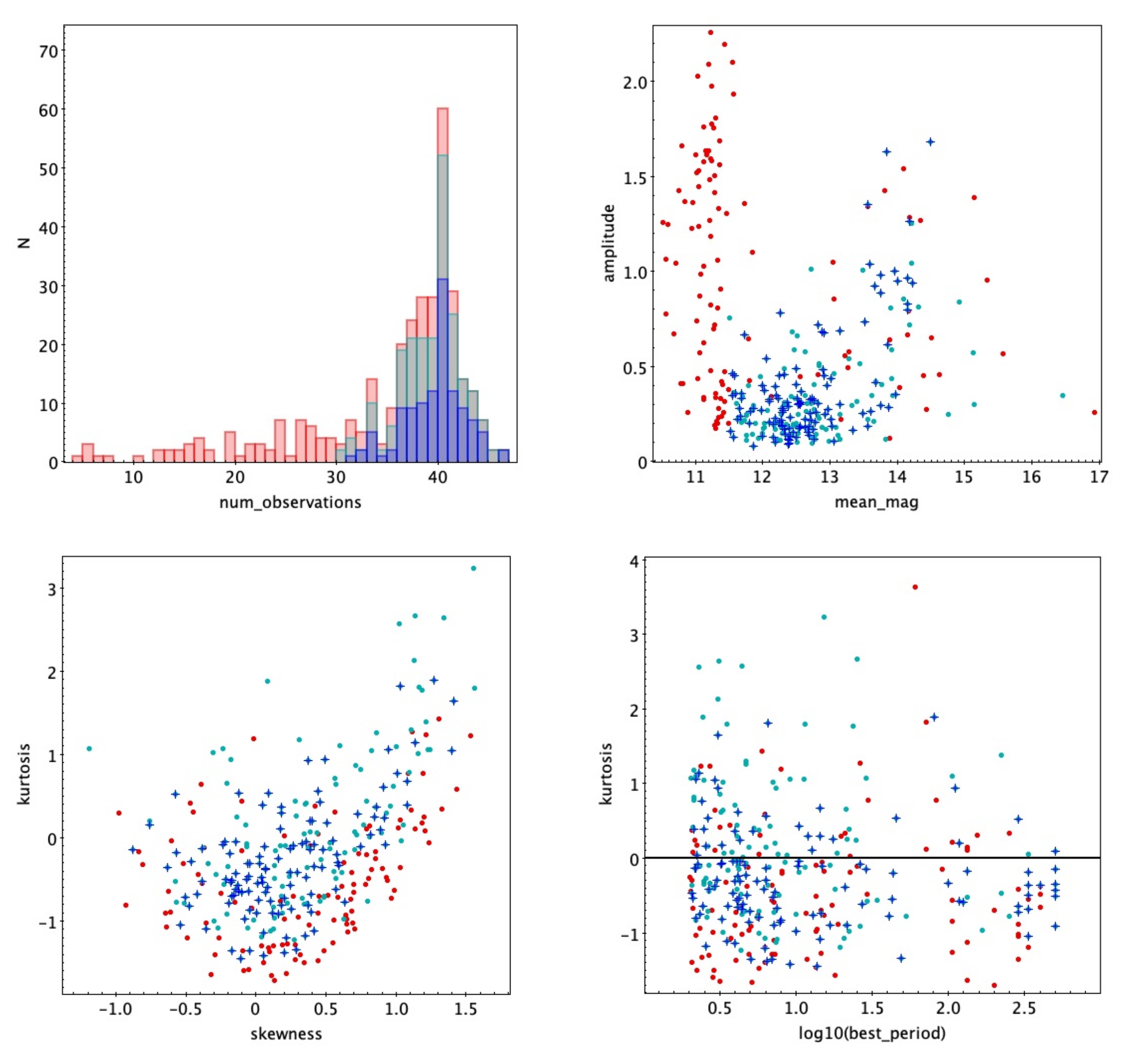}
 \caption{Results of the statistical analysis of the light curves for the most probable Carina members. Upper left: Histogram of the number of observations (epochs) per object. Red bars represent all sources, light green corresponds to reliable sources, and blue indicates periodic variables among the reliable ones. Upper right: Amplitude vs. mean $K_S$ magnitude diagram. Lower left: Skewness vs. kurtosis diagram. Lower right: Kurtosis vs. $\log$(best period) diagram. The color coding is the same for all three diagrams: red for all, light green for reliable, and red crosses marking periodic variables. These plots help identify the nature of the variability and highlight objects with non-Gaussian or burst-like behavior.}
\label{stats}
\end{center}
\end{figure*}

The catalog listed in Table\,\ref{catalog} is published in its entirety in the machine-readable format. A portion is shown here for guidance regarding its form and content.
The columns are as follows: Object ID; R.A. (deg); Decl. (deg); $K_S$ mean magnitude from VVVX; $K_S$ delta; $K_S$ amplitude;	$K_S$ variance; 	$K_S$ standard deviation; $K_S$ skewness;	$K_S$ kurtosis;	$K_S$ slope index;		$K_S$ GPV;		$K_S$ structure function; $K_S$ time span; $K_S$ best period; FAP; Number of observations;   Mass and errors (StarHorse); 	$T$eff(K) and errors (StarHorse);	[Fe/H] (dex) and errors (StarHorse); Reddening ($A_V$) and errors (StarHorse); and Classification flag: Em: Emissions in Br\,11, Abs: Absorptions in Br\,11, Var: Variable from VVVX; Mem: Probable member; Lit:  YSOs from known catalogs.

\begin{sidewaystable}\tabcolsep=2pt
\caption{Catalog of YSOs in Carina star forming region.}             
\centering    
\begin{tabular}{lccccccccccccccccccccccc}
\hline\hline 
  \multicolumn{1}{c}{OBJ\_ID} &
  \multicolumn{1}{c}{R.A.} &
  \multicolumn{1}{c}{Decl.} &
  \multicolumn{1}{c}{$K_S$\_mean} &
  \multicolumn{1}{c}{$K_S$\_delta} &
  \multicolumn{1}{c}{$K_S$\_amplitude} &
  \multicolumn{1}{c}{$K_S$\_variance} &
  \multicolumn{1}{c}{$K_S$\_std\_dev} &
  \multicolumn{1}{c}{$K_S$\_skewness} &
  \multicolumn{1}{c}{$K_S$\_kurtosis} &
  \multicolumn{1}{c}{$K_S$\_slope\_index} & 
  \multicolumn{1}{c}{$K_S$\_gpv} &\\
  \multicolumn{1}{c}{$K_S$\_structure\_function} &  
  \multicolumn{1}{c}{$K_S$\_time\_span} &
  \multicolumn{1}{c}{$K_S$\_best\_period} &
  \multicolumn{1}{c}{FAP} &
  \multicolumn{1}{c}{Num\_obser.} &
  \multicolumn{1}{c}{Class\_Flag} &
  \multicolumn{1}{c}{Flag\footnote{1-em, 2-ab, 3-lit, 4-var)\_1}} &
  \multicolumn{1}{c}{Lit.} &
   \multicolumn{1}{c}{Mass} &
 \multicolumn{1}{c}{$T$eff} & 
  \multicolumn{1}{c}{[Fe/H]} &
  \multicolumn{1}{c}{$A_V$}  \\
\hline
2M10345199-5923193	&	158.716663	&	-59.388702	&	13.766	&	0.554	&	0.406	&	0.02	&	0.142	&	0.489	&	-0.717	&	0.073	&	0.678	&\\
0.21	&	1785.988	&	5.699	&	0.362	&	40	&	abs,var	&	2	&		&	3.03	$\pm$	0.44	&	6404.0$\pm$184.7	&	-0.25$\pm$0.20	&	4.49$\pm$0.59	\\
&&&&&&&&&& \\[-3pt]
 2M10351877-5946276	&	158.828216	&	-59.774361	&	11.554	&	2.229	&	2.103	&	0.728	&	0.853	&	0.098	&	-1.363	&	0.021	&	0.496	&	\\
 1.025	&	1102.003	&	4.028	&	0.002	&	7	&	var,mem	&	4	&		&				&				&				&				\\
   &&&&&&&&&& \\[-3pt]
2M10352633-5936027	&	158.859734	&	-59.600754	&	12.216	&	0.163	&	0.116	&	0.001	&	0.039	&	0.364	&	-0.358	&	0.09	&	0.764	&	\\
0.05	&	1492.028	&	6.534	&	0.317	&	38	&	var,mem	&	4	&		&	1.66	$\pm$	0.19	&	4910.5$\pm$17.5	&	-0.13$\pm$0.02	&	3.47$\pm$0.34	\\
   &&&&&&&&&& \\[-3pt]
 2M10353239-5935204	&	158.884990	&	-59.589016	&	12.514	&	0.444	&	0.33	&	0.012	&	0.108	&	0.748	&	-0.112	&	0.123	&	0.81	&	\\
 0.137	&	1492.028	&	10.287	&	0.125	&	38	&	var,mem	&	3	&	M19,M23	&	1.33$\pm$0.14	&	4623.5$\pm$20.6	&	0.02$\pm$0.02	&	3.20$\pm$0.28	\\
   &&&&&&&&&& \\[-3pt]
 2M10354206-5937449	&	158.925290	&	-59.629150	&	13.009	&	0.31	&	0.221	&	0.005	&	0.071	&	0.36	&	-0.045	&	0.289	&	0.488	&	\\
 0.116	&	1492.028	&	4.305	&	0.018	&	36	&	var,mem	&	4	&		&	2.17$\pm$0.32	&	5034.2$\pm$51.0	&	-0.13$\pm$0.03	&	3.72$\pm$0.45	\\
   &&&&&&&&&& \\[-3pt]
 \multicolumn{11}{l}{... and 647 more stars.} \\
\hline   
\end{tabular}
\label{catalog}
\end{sidewaystable}



\section{Summary and conclusions}
In this study, we analyze 766 spectra in the Carina region obtained by APOGEE-2 medium-resolution $H$-band spectroscopy. To classify them, we applied an unsupervised K-means clustering algorithm, which separated the  sample into three major groups: "Emission-line YSOs" (with Br\,11 in emission), "Absorption-line YSOs" (with Br\,11 in absorption), and an "Unclassified" group. These groups are subsequently validated by manual inspection and comparison with existing catalogs of YSOs in the Carina Nebula.

Based on PSF photometry of more than 6.35 million sources across 6 VVVX tiles around Carina, and applying 2 variability indices ($\Delta K_{\rm S}$ and $\eta$), we identified 606 candidate variable stars. Among these, 143 objects originally from the "Unclassified" group have been reclassified and incorporated into the "Literature" group under the new label "Literature/Variable". 

Our final catalog of confirmed YSOs in the Carina star forming region contains 652 objects.

Proper motion and distance constraints derived from Gaia DR3 identified 415 probable Carina members, with mean distance estimates consistent with previous works.
The temperature, mass, and metallicity distributions were analyzed independently using spectroscopic measurements from the StarHorse database and Gaia DR3. Both the temperature and metallicity distributions are consistent with those typically observed for YSOs, with the majority of sources clustering around solar metallicity and effective temperatures in the range of 4000–6000\,K. Only 8 stars in the sample exhibit masses greater than 8\(M_\odot\), and 15 exceed 4\(M_\odot\). This indicates that the Carina Nebula is forming relatively few massive YSOs, suggesting a limited ongoing massive star formation in the current epoch, highlighting possible evolutionary implications or feedback-induced quiescence.

 The statistical characterization of YSO variability demonstrated that most Carina members (78\%) exhibit variability patterns. Of these, 134 stars are classified by our semisupervised K-means clustering algorithm  as "Emission-line YSOs", suggesting that they are active acceptors or at least show some undergoing episodic accretion processes. 
 
 This new dataset represents an ideal foundation for training machine learning algorithms to robustly classify YSOs and predict their observational signatures across diverse star-forming regions. Future work will expand on high-amplitude variable YSOs exhibiting unique spectral signatures, with targeted follow-up observations to better explain physical mechanisms underlying variability.

\begin{acknowledgements}
We are grateful to the anonymous reviewer for their helpful comments and valuable suggestions, which have helped improve the manuscript.  
This work was funded by ANID, Millennium Science Initiative, AIM23-0001. J.B. and R.K. acknowledge support from Fondecyt Regular 1240249. R.K.S. acknowledges support from CNPq/Brazil through projects 308298/2022-5 and 421034/2023-8. D.M. acknowledges support from the Center for Astrophysics and Associated Technologies CATA by the ANID BASAL projects ACE210002 and FB210003, and by Fondecyt Project No. 1220724. P.W.L acknowledges support by grant ST/Y000846/1 of the UK Science and Technology Facilities Council.  This work was supported by CASSACA and ANID through the China-Chile Joint Research Funding CCJRF2301.

Funding for the Sloan Digital Sky 
Survey IV (SDSS-IV)has been provided by the 
Alfred P. Sloan Foundation, the US 
Department of Energy Office of 
Science, and participating 
institutions. 
SDSS-IV acknowledges support and 
resources from the Center for High 
Performance Computing  at the 
University of Utah. The SDSS 
website is www.sdss4.org.
SDSS-IV is managed by the 
Astrophysical Research Consortium 
for the Participating Institutions 
of the SDSS Collaboration including 
the Brazilian Participation Group, 
the Carnegie Institution for Science, 
Carnegie Mellon University, Center for 
Astrophysics | Harvard \& 
Smithsonian, the Chilean Participation 
Group, the French Participation Group, 
Instituto de Astrof\'isica de 
Canarias, The Johns Hopkins 
University, Kavli Institute for the 
Physics and Mathematics of the 
Universe (IPMU)/University of 
Tokyo, the Korean Participation Group, 
Lawrence Berkeley National Laboratory, 
Leibniz Institut f\"ur Astrophysik 
Potsdam (AIP),  Max-Planck-Institut 
f\"ur Astronomie (MPIA Heidelberg), 
Max-Planck-Institut f\"ur 
Astrophysik (MPA Garching), 
Max-Planck-Institut f\"ur 
Extraterrestrische Physik (MPE), 
National Astronomical Observatories of 
China, New Mexico State University, 
New York University, University of 
Notre Dame, Observat\'ario 
Nacional/MCTI, The Ohio State 
University, Pennsylvania State 
University, Shanghai 
Astronomical Observatory, United 
Kingdom Participation Group, 
Universidad Nacional Aut\'onoma 
de M\'exico, University of Arizona, 
University of Colorado Boulder, 
University of Oxford, University of 
Portsmouth, University of Utah, 
University of Virginia, University 
of Washington, University of 
Wisconsin, Vanderbilt University, 
and Yale University.

This work has made use of data from the European Space Agency (ESA) mission
{\it Gaia} (\url{https://www.cosmos.esa.int/gaia}), processed by the {\it Gaia}
Data Processing and Analysis Consortium (DPAC;
\url{https://www.cosmos.esa.int/web/gaia/dpac/consortium}). Funding for the DPAC
has been provided by national institutions, in particular the institutions
participating in the {\it Gaia} Multilateral Agreement.
\end{acknowledgements}

\newpage
\bibliography{carina}{}
\bibliographystyle{aasjournalv7}

 \end{document}